\documentclass[showpacs,preprint,superscriptaddress,amsmath,amssymb]{revtex4-1}
\usepackage{txfonts}
\usepackage{graphicx}
\usepackage{dcolumn}
\usepackage{bm}
\usepackage{epsfig}
\usepackage{booktabs}
\usepackage{subfigure}
\usepackage{graphics}
\usepackage{amssymb}
\usepackage{amsmath}
\usepackage{array}
\usepackage{color}
\usepackage{tabularx}
\usepackage{multirow}

\usepackage{CJK}
\usepackage{xcolor}
\usepackage{algorithm}
\usepackage{algpseudocode}

\begin{document}

\title{Rectangular multiple-relaxation-time lattice Boltzmann method for the Navier-Stokes and nonlinear convection-diffusion equations: General equilibrium and some important issues}

\author{Zhenhua Chai}\email{hustczh@hust.edu.cn}
\affiliation{Institute of Interdisciplinary Research for Mathematics and Applied Science, School of Mathematics and Statistics, Huazhong
University of Science and Technology, Wuhan 430074, China}
\affiliation{Hubei Key Laboratory of Engineering Modeling and
Scientific Computing, Huazhong University of Science and Technology,
Wuhan 430074, China}

\author{Xiaolei Yuan}\email{yuan_xiaolei2009@163.com}
\affiliation{College of Mathematics and Information Science, Hebei University, Baoding 071002, China}

\author{Baochang Shi}
\email[Corresponding author] {(shibc@hust.edu.cn)}
\affiliation{Institute of Interdisciplinary Research for Mathematics and Applied Science, School of Mathematics and Statistics, Huazhong
University of Science and Technology, Wuhan 430074, China}
\affiliation{Hubei Key Laboratory of Engineering Modeling and
Scientific Computing, Huazhong University of Science and Technology,
Wuhan 430074, China}

\date{\today}

\begin{abstract}
In this paper, we develop a general rectangular multiple-relaxation-time lattice Boltzmann (RMRT-LB) method for the Navier-Stokes equations (NSEs) and nonlinear convection-diffusion equation (NCDE) by extending our recent unified framework of MRT-LB method [Z.H. Chai and B.C. Shi, Phys. Rev. E 102, 023306 (2020)], where a rectangular equilibrium distribution function (REDF) [J.H. Lu et al, Phil. Trans. R. Soc. A 369, 2311-2319 (2011)] on a rectangular lattice is utilized. 
 Due to the anisotropy of discrete velocities on a rectangular lattice, the third-order moment of REDF is inconsistent with that of popular LB method, and thus the previous unified framework of MRT-LB method cannot be directly applied to the NSEs using the REDF on the rectangular rD$d$Q$q$ ($q$ discrete velocities in $d$-dimensional space, $d \geq1$) lattice. The macroscopic NSEs can be recovered from the RMRT-LB method through the direct Taylor expansion method by properly selecting the relaxation sub-matrix $\mathbf{S}_2$ which is related to kinetic viscosity and bulk viscosity. While the rectangular lattice does not lead to the change of the zero-th, first and second-order moments of REDF, thus the unified framework of MRT-LB method can be directly applied to the NCDE. It should be noted that the RMRT-LB model for NSEs can be derived on the rD$d$Q$q$ lattice, including rD2Q9, rD3Q19, and rD3Q27 lattices, while there are no rectangular D3Q13 and D3Q15 lattices within this framework of RMRT-LB method. Thanks to the block-lower triangular-relaxation matrix introduced in the unified framework, the RMRT-LB versions (if existed) of the previous MRT-LB models can be obtained, including those based on raw (natural)-moment, central-moment, Hermite-moment, and central Hermite-moment, respectively. It is also found that when the standard or regular lattice is used, and the sound speed is taken as an adjustable parameter, the present RMRT-LB method becomes a new class of MRT-LB method for the NSEs and NCDE, and the commonly used MRT-LB models on the D$d$Q$q$ lattice are only its special cases. 
\end{abstract}


\maketitle

\section{Introduction}

The lattice Boltzmann (LB) method, as an effective mesoscopic numerical approach based on the kinetic theory, has attained increasing attention, and also gained a great success in the modeling and simulation of the complex fluid flows, heat and mass transfer described by the Navier-Stokes equations (NSEs) and nonlinear convection-diffusion equations (NCDEs) \cite{Chen1998,Succi2001,Guo2013,Kruger2017}.
Although many different LB models have been developed in the past three decades, most of them can be viewed as the special forms of the multiple-relaxation-time lattice Boltzmann (MRT-LB) model \cite{dHumieres2002}, except those with nonlinear collision operators (e.g., the cumulant LB model  \cite{Geier2015}). The MRT-LB model uses a linear collision operator, that is, a collision matrix with multiple relaxation parameters, which extends the commonly used single-relaxation-time LB (SRT-LB) model or lattice Bhatnagar-Gross-Krook (LBGK) model so that the numerical stability and accuracy can be improved by adjusting the free relaxation parameters.  Actually, some previous works have shown that the MRT model is superior to the SRT model in terms of accuracy and stability, but only slightly inferior to the SRT model in the computational efficiency \cite{dHumieres2002,Pan2006,Luo2011}.
As we know, in the standard LB method for fluid flows, its evolution is carried out in a highly symmetric lattice space through two steps: collision and propagation. This makes the computational grid and lattice space coupled, resulting in that the standard LB model is generally only implemented on a uniform grid, e.g., the square lattice in two-dimensional space and the cubic lattice in three-dimension lattice. It should be noted that this restriction is caused by the requirement on the isotropy of the standard discrete velocity set, and also limits the applications of the standard LB model.

There are mainly four ways to design the LB method on a non-uniform grid:

(1) The method in which the computational grid and lattice space are decoupled \cite{Guo2013a,GuoZhao2003}.

This method combines the LB approach with some numerical scheme (e.g., the finite volume scheme, finite difference scheme, and finite element scheme) for the discrete velocity Boltzmann equation (DVBE), and thus both of the non-uniform grid and uniform (isotropic) lattice can be used in this kind of method.

(2) The interpolation-supplemented LB method \cite{He1996,Niu2003}.

In this method, spatial and temporal interpolation schemes are used to map the information on an inherent lattice grid to a computational grid where the hydrodynamic variables are obtained. Such implementations preserve the main advantages of standard LBM, and are more flexible in selecting the computational grid. However, to retain the second-order accuracy of the LB method, a second-order or high-order interpolation must be used, which results in additional computational cost. In addition, some undesirable errors and numerical dissipation may be introduced by the interpolation, which in turn brings numerical stability concerns.

(3) The LB method with local grid refinement \cite{Filippova1998,Guo2003DDLB,Lagrava2012}.

In this method, the whole computational domain is divided into several sub-domains, and the standard LB model is used in each sub-domain which has its own uniform lattice. Two adjacent sub-domains with different lattices may share a common ghost boundary, and the distribution functions (DFs) and/or physical quantities on the ghost boundary need to be determined by some suitable interpolation schemes from both sides of the ghost boundary. Generally, the domain-decomposition technique can be utilized in dividing the domain, and each sub-domain can use its own coordinates system and regular lattice \cite{Guo2003DDLB}. However, the shortcomings caused by interpolation are also remained in this method.

(4) The LB method on the rectangular lattice.

Unlike the above methods, this method is expected to be a natural and direct extension of the standard LB method on a uniform lattice, and also keeps the \emph{collision-propagation} characteristics of the latter. Compared to the rectangular LB (RLB) model for NCDE, the development of the RLB model for NSEs is not well due to the inconsistency between the anisotropy of discrete veolcitites caused by the rectangular lattice and the isotropic viscosity of the NSEs. 
 In fact, when the SRT collision operator is used, the degrees of freedom of the standard D2Q9 lattice and D3Q19 lattice, even D3Q27 lattice are not enough to remove the anisotropy \cite{Hegeler2013}. As far as we know, Koelman \cite{Koelman1991} first proposed a SRT-LB model on a rectangular lattice where a low Mach-number expansion of the Maxwell-Boltzmann distribution is used to obtain the equilibrium distribution function. However, this model fails to recover the correct NSEs with isotropic viscosity when the grid aspect ratio is different from $1$. Then, to derive the correct NSEs, several modified rectangular SRT-LB models were proposed. For example, Hegeler et al. \cite{Hegeler2013} developed a rectangular SRT-LB model that can derive the correct macroscopic equations by adding discrete velocity directions to increase degrees of freedom. Peng et al. \cite{PengGuoWang2019} and Saadat et al. \cite{Saadat2021} respectively proposed another SRT-LB model on a rectangular grid that can reproduce the correct NSEs by introducing an extended equilibrium distribution function. Besides, Wang et al. \cite{Wang2019} constructed a SRT-LB model through including some artificial counteracting forcing terms, which are used to remove the anisotropy caused by the rectangular lattice. We note that these SRT-LB models mentioned above can be regarded as some modifications to the standard SRT-LB model by introducing additional velocity directions, extended equilibrium distribution function or force terms, which may bring more computational cost. Additionally, it is also not flexible enough to adjust the shear and bulk viscosities independently.

 Considering the fact that compared to the SRT-LB model, the MRT-LB model can provide additional degrees of freedom, another feasible approach is to use the MRT collision model to remove this inconsistency. Under the framework of MRT-LB method, Bouzidi et al. \cite{Bouzidi2001}  constructed the first MRT-LB model on a two-dimensional rectangular grid. Although this model can obtain the correct results in some numerical tests, the anisotropy problem cannot be overcome completely when the grid aspect ratio does not equal to one \cite{ZongPengGuo2016}. A similar attempt has been made by Zhou \cite{ZhouJG2012}, however this MRT model on the rectangular lattice cannot correctly recover the NSEs. Through an inverse design analysis based on the Chapman-Enskog expansion, Zong et al. \cite{ZongPengGuo2016} proposed another MRT-LB model on the D2Q9 rectangular grid, in which an additional adjustable parameter that governs the relative orientation in the energy-normal stress subspace is introduced. By adjusting this parameter, this model can correctly recover the macroscopic equations. However, this model is complicated, and is difficult to be extended to other lattice structures. After that, Peng et al. \cite{PengMinGuo2016} designed an alternative MRT-LB model on a rectangular grid by incorporating stress components into the equilibrium moments to remove the anisotropy in the stress tensor. Based on the work of  Peng et al. \cite{PengMinGuo2016}, Wang et al. \cite{WangMinPengGuo2019} developed a D3Q19 MRT-LBM model on a general cuboid grid. However, these two models require a relatively complex quasi-equilibrium collision step. In addition, Yahia et al. \cite{Yahia2021a} developed a rectangular central-moment MRT-LBM based on a non-orthogonal moment basis, and then extended it to three-dimensional central-moment LBM on a cuboid lattice \cite{Yahia2021b}. The equilibrium to which the central moments relax under collision in this approach is obtained from those corresponding to the continuous Maxwell distribution. These two models involve a relatively complicated computation of the equilibrium moments.

Although the correct macroscopic equations can be derived by introducing artificial source terms, additional adjustable parameter or extended equilibrium distribution function, these treatments are not necessary. Actually, the correct NSEs can be obtained by exploiting the properties of the rectangular MRT model without adding any additional terms. Based on this idea, Zecevic et al. \cite{Zecevic2020} presented a RLB method on two and three-dimensional lattices. They adopted a different set of basis vectors that allows hydrodynamic behaviour to be restored by adjusting relaxation parameters alone. We note that the RMRT model of Zecevic et al. \cite{Zecevic2020} still has some certain limitations. For instance, as stated in Ref. \cite{Zecevic2020}, there is no the orthogonal transformation matrix for rD3Q19, thus the RMRT-LB model with rD3Q19 lattice cannot be given \cite{Zecevic2020}. However, the RMRT-LB model with rD3Q19 lattice actually exists. In addition, the equilibrium distribution function and the bulk viscosity in the NSEs are also not considered in Ref. \cite{Zecevic2020}.

In this work, we intend to construct a general RMRT-LB model for the Navier-Stokes and nonlinear convection-diffusion equations by extending the recent unified framework of standard MRT-LB method \cite{Chai2020}, and derive a general rectangular equilibrium distribution function proposed by Lu et al. \cite{LuChaiShi2011} in detail. Here we would like to point out that in most of the RMRT-LB models, the collision process is carried out in the moment space rather than the velocity (distribution function) space, thus the analysis method (e.g., the Chapman-Enskog expansion) is very complicated, and often depends on the specified lattice structure or discrete velocity set. In view of this,  we will follow the previous work \cite{Chai2020} to conduct the modeling and analysis in the distribution function space, and show that the present model is more general and more natural. For simplicity, we only give some basic elements in the implementation of the RMRT-LB method, including the velocity moments, the weights of different lattice models (e.g., rD2Q9, rD3Q19, and rD3Q27 lattices), the equilibrium, auxiliary and source distribution functions. Through the direct Taylor expansion method, a generalized NSEs with a viscosity tensor can be correctly recovered through properly selecting the relaxation sub-matrix $\mathbf{S}_2$, which is related to kinetic viscosity and bulk viscosity. In addition, the present RMRT-LB model would reduce to the unified framework of MRT-LB method \cite{Chai2020} with the grid aspect ratio being $1$ and sound speed $c_s^2=c^2/3$. If the sound speed is taken as an adjustable parameter and the standard lattice is used, the present RMRT-LB method can be considered as a new class of MRT-LB method for the NSEs and CDEs. We would also like to point out that the present RMRT-LB model is a natural extension to the previous work  \cite{Chai2020}, and it does not introduce any assumptions or additional computational steps. Additionally, compared to the available RLB models, the present one is very simple and easy to implement.

The rest of this paper is organized as follows. In Sec. \uppercase\expandafter{\romannumeral2}, the rectangular multiple-relaxation-time lattice Boltzmann method is presented, and the general equilibrium, auxiliary and source distribution functions of RMRT-LB method on the rectangular lattice is given in Sec. \uppercase\expandafter{\romannumeral3}. Then the direct Taylor expansion analysis of present RMRT-LB method is conducted to recover the macroscopic equations in Sec. \uppercase\expandafter{\romannumeral4}. After that, we present the structure of collision matrix and some special cases of the RMRT-LB method in Sec. \uppercase\expandafter{\romannumeral5}. Finally, some conclusions are summarized in Sec. \uppercase\expandafter{\romannumeral6}.

\section{Rectangular Multiple-relaxation-time lattice Boltzmann method}

The evolution equation of the RMRT-LB method with the rectangular D$d$Q$q$ (rD$d$Q$q$) lattice model has the same form as that of the MRT-LB method  \cite{Chai2020}:
\begin{equation}\label{eq:2-1}
 f_j(\mathbf{x}+\mathbf{c}_j \Delta t,t+\Delta t)=f_j(\mathbf{x}, t)-\mathbf{\Lambda}_{jk} f_k^{ne}(\mathbf{x}, t)+\Delta t \big[G_j(\mathbf{x},t)+F_j(\mathbf{x}, t)+\frac{\Delta t}{2}\bar{D}_j F_j(\mathbf{x}, t)\big],
\end{equation}
where $f_j(\mathbf{x}, t)$ is the distribution function (DF) at position $\mathbf{x}$ in $d$-dimensional space and time $t$ along the velocity $\mathbf{c}_j$, $f_j^{ne}(\mathbf{x}, t)=f_j(\mathbf{x}, t)-f_j^{eq}(\mathbf{x}, t)$ is the nonequilibrium distribution function (NEDF), and $f_j^{eq}(\mathbf{x}, t)$ is the equilibrium distribution function (EDF). $F_j(\mathbf{x}, t)$ is the DF of a source or forcing term, $G_j(\mathbf{x}, t)$ is the \textit{auxiliary} DF, and $\mathbf{\Lambda}=(\mathbf{\Lambda}_{jk})$ is a $q \times q$ invertible collision matrix. $\Delta t$ is the time step, $\bar{D}_j=\partial_t +\gamma \mathbf{c}_j\cdot \nabla$ with $\gamma =1$ for NSEs and $\gamma \in \{0,1\}$ for NCDE. In the evolution equation (\ref{eq:2-1}), the key elements, $\mathbf{c}_j, f_{j}^{eq}, F_j, G_j$ and $\mathbf{\Lambda}$, must be given properly.

The unknown macroscopic conserved variable(s) $\phi(\mathbf{x},t)$ for NCDE, or $\rho(\mathbf{x},t)$ and $\mathbf{u}(\mathbf{x},t)$ for NSEs, are computed by

\begin{subequations}\label{eq:2-4-0}
\begin{equation}
\phi(\mathbf{x}, t)=\sum_j f_j(\mathbf{x}, t),
\end{equation}
\begin{equation}
\rho(\mathbf{x}, t)=\sum_j f_j(\mathbf{x}, t),
\mathbf{u}(\mathbf{x}, t)=\sum_j \mathbf{c}_j f_j(\mathbf{x}, t)/\rho(\mathbf{x}, t).
\end{equation}
\end{subequations}

The evolution equation (\ref{eq:2-1}) can be divided into two sub-steps, i.e., collision and propagation,
\begin{subequations}\label{eq:2-2}
\begin{equation}
\textbf{Collison:}\ \  \tilde{f}_j(\mathbf{x}, t)=f_j(\mathbf{x}, t)-\mathbf{\Lambda}_{jk} f_k^{ne}(\mathbf{x}, t)+\Delta t \big[G_j(\mathbf{x},t)+F_j(\mathbf{x}, t)+\frac{\Delta t}{2}\bar{D}_j F_j(\mathbf{x}, t)\big],  \\
\end{equation}
\begin{equation}
\textbf{Propagation:}\ \  f_j(\mathbf{x}+\mathbf{c}_j \Delta t,t+\Delta t)=\tilde{f}_j(\mathbf{x}, t), \ \ \ \ \ \ \ \ \ \ \ \ \ \ \ \ \ \ \ \ \ \ \ \ \ \ \ \ \ \ \ \ \ \ \ \ \ \ \ \ \ \ \ \ \ \ \ \ \ \ \ \ \ \ \ \ \ \ \ \
\end{equation}
\end{subequations}
where $\tilde{f}_j(\mathbf{x}, t)$ is the post-collision distribution function.

As pointed out in Ref.  \cite{Chai2020}, in almost all of the MRT-LB models, the collision process in Eq. (\ref{eq:2-1}) is carried out in the moment space, and the analysis method (e.g., the popular Chapman-Enskog analysis) is usually more complicated and often depends on the specified lattice structure or discrete velocity set  \cite{dHumieres1992,Guo2013,Chai2012,Yoshida2010,Chai2016a}. In this work, we will extend the unified framework of MRT-LB method \cite{Chai2020} to that of RMRT-LB method.

In the implementation of the RMRT-LB method, there are two popular schemes that can be used to discretize $\bar{D}_j F_j(\mathbf{x}, t)$ in the last term on the right hand side of Eq. (\ref{eq:2-1}). Actually, if $\gamma=0$, the first-order explicit difference scheme $\partial_{t} F_j(\mathbf{x}, t)=[F_j(\mathbf{x}, t)-F_j(\mathbf{x}, t-\Delta t)]/\Delta t$ is adopted for NCDEs  \cite{Shi2009,Chai2016a}. While for the case of $\gamma=1$, we can use the first-order implicit difference scheme $(\partial_{t}+\mathbf{c}_{j}\cdot\nabla) F_j(\mathbf{x}, t)=[F_j(\mathbf{x}+\mathbf{c}_{j}\Delta t, t+\Delta t)-F_j(\mathbf{x}, t)]/\Delta t$ for both NCDE and NSEs, and take the  transform $\bar{f}_j=f_j-\frac{\Delta t}{2}F_j$ as in Refs. \cite{He1998, Du2006,Chai2016a}, then Eq. (\ref{eq:2-1}) becomes
\begin{equation}\label{eq:2-3}
 \bar{f}_j(\mathbf{x}+\mathbf{c}_j \Delta t,t+\Delta t)=\bar{f}_j(\mathbf{x}, t)-\mathbf{\Lambda}_{jk} \bar{f}_k^{ne}(\mathbf{x}, t)+\Delta t \big[G_j(\mathbf{x},t)+(\delta_{jk}-\mathbf{\Lambda}_{jk}/2)F_k(\mathbf{x}, t)\big],
\end{equation}
where $\bar{f}_j^{ne}(\mathbf{x}, t)=\bar{f}_j(\mathbf{x}, t)-f_j^{eq}(\mathbf{x}, t)$. Additionally, we also have the following relations,
\begin{subequations}\label{eq:2-4}
\begin{equation}
\sum_j f_j(\mathbf{x}, t)= \sum_j\bar{f}_j(\mathbf{x}, t)+\frac{\Delta t}{2}\sum_j F_j(\mathbf{x}, t),
\end{equation}
\begin{equation}
\sum_j \mathbf{c}_j f_j(\mathbf{x}, t)= \sum_j \mathbf{c}_j\bar{f}_j(\mathbf{x}, t)+\frac{\Delta t}{2}\sum_j \mathbf{c}_j F_j(\mathbf{x}, t).
\end{equation}
\end{subequations}

It follows from Eqs. (\ref{eq:2-4-0}) and (\ref{eq:2-4}) that for NCDE,

\begin{equation}\label{eq:2-4-1}
\phi(\mathbf{x}, t)= \sum_j\bar{f}_j(\mathbf{x}, t)+\frac{\Delta t}{2}\sum_j F_j(\mathbf{x}, t),
\end{equation}
or for NSEs,
\begin{subequations}\label{eq:2-4-2}
\begin{equation}
\rho(\mathbf{x}, t)= \sum_j\bar{f}_j(\mathbf{x}, t)+\frac{\Delta t}{2}\sum_j F_j(\mathbf{x}, t),
\end{equation}
\begin{equation}
\mathbf{u}(\mathbf{x}, t)= \big[\sum_j \mathbf{c}_j\bar{f}_j(\mathbf{x}, t)+\frac{\Delta t}{2}\sum_j \mathbf{c}_j F_j(\mathbf{x}, t)\big]/\rho(\mathbf{x}, t).
\end{equation}
\end{subequations}

\section{General equilibrium, auxiliary and source distribution functions of RMRT-LB method on rectangular lattice}

As we know, to recover the NSEs (\ref{eq:NSE}) and NCDE (\ref{eq:NCDE}) from Eq. (\ref{eq:2-1}) correctly, some proper moments of the equilibrium, auxiliary and source distribution functions should be given correctly.

We first consider the rD$d$Q$q$ lattice models. To simplify the following analysis, we introduce
 $c_\alpha=\Delta x_{\alpha}/\Delta t$ ($\alpha=1, 2, \ldots, d$) in $d$-dimensional space with $\Delta x_{\alpha}$ being the spacing step in $\alpha$ axis. In this case, the discrete velocities and weight coefficients in the common rD$d$Q$q$ lattice models can be given as follows.

rD2Q9 lattice:

\begin{equation}\label{eq:3-1}
\begin{split}
&\{\mathbf{c}_j,0\leq j \leq 8\}=
\left(
\begin{array}{ccccccccc}
    0 & c_1 &   0 & -c_1  &    0 & c_1 & -c_1 & -c_1 &  c_1 \\
    0 &   0 & c_2 &    0  & -c_2 & c_2 &  c_2 & -c_2 & -c_2 \\
\end{array}
\right),\\
&\omega_j\geq 0, \omega_1=\omega_3,\omega_2=\omega_4,\omega_5=\omega_6=\omega_7=\omega_8, \omega_0=1-\sum_{j>0} \omega_j.
\end{split}
\end{equation}
There are two lattice models that can be obtained as the subsets of rD2Q9 lattice by setting some weights to zero and removing some velocities in the rD2Q9 lattice. The first one is rD2Q5I lattice, in which $\omega_j=0$ $(j>4)$ and the discrete velocities $\mathbf{c}_5$-$\mathbf{c}_8$ are removed. The second is the rD2Q5II lattice where $\omega_j=0$ $(1\leq j \leq 4)$  and the discrete velocities $\mathbf{c}_1$-$\mathbf{c}_4$ are removed.

rD3Q27 lattice:

\begin{subequations}\label{eq:3-2}
\begin{equation}\label{eq:3-2a}
\{\mathbf{c}_j,0\leq j \leq 6\}=
\left(
\begin{array}{ccccccc}
    0 & c_1 &   0 &   0 & -c_1  &    0 &    0 \\
    0 &   0 & c_2 &   0 &    0  & -c_2 &    0 \\
    0 &   0 &   0 & c_3 &    0  &    0 & -c_3 \\
\end{array}
\right),
\end{equation}
\begin{equation}\label{eq:3-2b}
\{\mathbf{c}_j,7\leq j \leq 18\}=
\left(
\begin{array}{cccccccccccc}
    c_1 & -c_1 & -c_1 &  c_1 & c_1 & -c_1 & -c_1 & c_1 &   0 &    0 &    0 &    0 \\
    c_2 &  c_2 & -c_2 & -c_2 &   0 &    0 &    0 &   0 & c_2 & -c_2 & -c_2 &  c_2 \\
      0 &    0 &    0 &    0 & c_3 & -c_3 & -c_3 & c_3 & c_3 &  c_3 & -c_3 & -c_3 \\
\end{array}
\right),
\end{equation}
\begin{equation}\label{eq:3-2c}
\{\mathbf{c}_j,19\leq j \leq 26\}=
\left(
\begin{array}{cccccccc}
    c_1 &  c_1 &  c_1 & -c_1 & -c_1 & -c_1 & -c_1 &  c_1     \\
    c_2 &  c_2 & -c_2 &  c_2 & -c_2 & -c_2 &  c_2 & -c_2     \\
    c_3 & -c_3 &  c_3 &  c_3 & -c_3 &  c_3 & -c_3 & -c_3     \\
\end{array}
\right),
\end{equation}
\begin{equation*}
\begin{split}
\omega_j\geq 0, \omega_1&=\omega_4,\omega_2=\omega_5,\omega_3=\omega_6,\omega_7=\omega_8=\omega_9=\omega_{10}, \omega_{11}=\omega_{12}=\omega_{13}=\omega_{14},\\
&\omega_{15}=\omega_{16}=\omega_{17}=\omega_{18},\omega_j=\omega_{19} (j>19), \omega_0=1-\sum_{j>0} \omega_j.
\end{split}
\end{equation*}
\end{subequations}

Similarly, we can obtain the following subsets of rD3Q27 lattice.

rD3Q19 lattice: $\omega_{19}=0$ and the discrete velocities $\mathbf{c}_j$ $(j \geq 19)$ are removed.

rD3Q15 lattice: $\omega_{7}=\omega_{11}=\omega_{15}=0$ and the discrete velocities $\mathbf{c}_j$ $(j=7-18)$ are removed.

rD3Q13 lattice: $\omega_1=\omega_2=\omega_3=\omega_{19}=0$ and the discrete velocities $\mathbf{c}_j$ $(j=1-6, 19-26)$ are removed.

rD3Q9 lattice: $\omega_j=0$ $(1\leq j \leq 18)$ and the discrete velocities $\mathbf{c}_j$ $(j=1-18)$ are removed.

rD3Q7 lattice: $\omega_j=0$ $(j>6)$ ) and the discrete velocities $\mathbf{c}_j$ $(j=7-26)$ are removed.

We note that the weights $\omega_j$ in the rD$d$Q$q$ lattice can be determined by the velocity moments. Let $\mathbf{\Delta}^{(m)}$ be the $m$-th order moment of velocity set $\{\mathbf{c}_j,0\leq j\leq q-1\}$,
\begin{equation}\label{eq:3-3-1}
\mathbf{\Delta}^{(m)}=\sum_j \omega_j \underbrace{ \mathbf{c}_j\mathbf{c}_j\cdots \mathbf{c}_j}_{m},
\end{equation}
then due to the lattice symmetry, all of the odd-order moments are zero, and the even-order moments $\mathbf{\Delta}^{(0)}$, $\mathbf{\Delta}^{(2)}$ and $\mathbf{\Delta}^{(4)}$ are used to determine the expression of EDF $f_j^{eq}$. Actually, to derive correct macroscopic equations, $\mathbf{\Delta}^{(0)}=\sum_j \omega_j=1$, $\mathbf{\Delta}^{(2)}$ and $\mathbf{\Delta}^{(4)}$ should be given properly.

It is a natural extension that $\mathbf{\Delta}^{(2)}$ is taken as  \cite{LuChaiShi2011}
\begin{equation}\label{eq:3-3-2}
\mathbf{\Delta}^{(2)}=\sum_j \omega_j\textbf{c}_j\textbf{c}_j= \mathbf{diag}(c_{s\alpha}^2),
\end{equation}
where the lattice speed $c_{s\alpha}$ is taken as a parameter along $\alpha$-axis.

For rD2Q9 lattice (\ref{eq:3-1}), it is easy to obtain the following results,
\begin{equation}\label{eq:3-3-3}
\mathbf{\Delta}^{(2)}_{\alpha\alpha}=\sum_j \omega_j\textbf{c}_{j\alpha}^2=2c_{\alpha}^2(\omega_{\alpha}+2\omega_5)= c_{s\alpha}^2; \mathbf{\Delta}^{(2)}_{\alpha\beta}=0,\alpha \neq \beta,
\end{equation}
\begin{equation}\label{eq:3-3-4}
\mathbf{\Delta}^{(4)}_{\alpha\alpha\alpha\alpha}=c_{\alpha}^2\mathbf{\Delta}^{(2)}_{\alpha\alpha}= c_{\alpha}^2 c_{s\alpha}^2;\\ \mathbf{\Delta}^{(4)}_{\alpha\alpha\beta\beta}=4c_{\alpha}^2 c_{\beta}^2 \omega_5,\alpha\neq\beta;\\  \mathbf{\Delta}^{(4)}_{\alpha\beta\gamma\theta}=0,else.
\end{equation}
From Eqs. (\ref{eq:3-3-3}) and (\ref{eq:3-1}), we have $0<2 \omega_{\alpha}+4\omega_5=c_{s\alpha}^2/c_{\alpha}^2<1$, and then $0 < \omega_5<c_{s\alpha}^2/(4c_{\alpha}^2)$ $(\alpha=1,2).$
Compared to standard lattice, if we take $\mathbf{\Delta}^{(4)}_{\alpha\alpha\beta\beta}=\mathbf{\Delta}^{(2)}_{\alpha\alpha}\mathbf{\Delta}^{(2)}_{\beta\beta}=c_{s\alpha}^2 c_{s\beta}^2 (\alpha\neq \beta)$ in Eq. (\ref{eq:3-3-4}), one can derive
\begin{equation}\label{eq:3-3-5}
\omega_5=\frac{c_{s1}^2 c_{s2}^2}{4c_1^2 c_2^2},\omega_{\alpha}=\frac{c_{s\alpha}^2}{2c_{\alpha}^2}-2\omega_5,\alpha=1,2.
\end{equation}

It should be noted that the constraint $\mathbf{\Delta}^{(4)}_{\alpha\alpha\beta\beta}=\mathbf{\Delta}^{(2)}_{\alpha\alpha}\mathbf{\Delta}^{(2)}_{\beta\beta}$ $(\alpha\neq \beta)$ is equivalent to the weighted orthogonality condition of $Q_{i\alpha\alpha}$ and $Q_{i\beta\beta}$  $(\alpha\neq \beta)$, $\mathbf Q_j=\mathbf c_j \mathbf c_j - \mathbf \Delta^{(2)}$ is the second discrete Hermite polynomial weighted orthogonal to $\mathbf c^0_j$ and $\mathbf c_j$.

Similarly, for rD3Q27 lattice (\ref{eq:3-2}), we have
\begin{subequations}\label{eq:3-3-6}
\begin{equation}
\mathbf{\Delta}^{(2)}_{11}=2c_1^2(\omega_1+2(\omega_7+\omega_{11})+4\omega_{19})= c_{s1}^2,\ \ \
\end{equation}
\begin{equation}
\mathbf{\Delta}^{(2)}_{22}=2c_2^2(\omega_2+2(\omega_7+\omega_{15})+4\omega_{19})= c_{s2}^2,\ \ \
\end{equation}
\begin{equation}
\mathbf{\Delta}^{(2)}_{33}=2c_3^2(\omega_3+2(\omega_{11}+\omega_{15})+4\omega_{19})= c_{s3}^2,\ \ \
\end{equation}
\end{subequations}
\begin{subequations}\label{eq:3-3-7}
\begin{equation}
\mathbf{\Delta}^{(4)}_{1122}=4c_1^2 c_2^2(\omega_7+2\omega_{19})= c_{s1}^2 c_{s2}^2,\ \ \
\end{equation}
\begin{equation}
\mathbf{\Delta}^{(4)}_{1133}=4c_1^2 c_3^2(\omega_{11}+2\omega_{19})= c_{s1}^2 c_{s3}^2,\ \ \
\end{equation}
\begin{equation}
\mathbf{\Delta}^{(4)}_{2233}=4c_2^2 c_3^2(\omega_{15}+2\omega_{19})= c_{s2}^2 c_{s3}^2,\ \ \
\end{equation}
\begin{equation}
\mathbf{\Delta}^{(4)}_{\alpha\alpha\alpha\alpha}=c_{\alpha}^2\mathbf{\Delta}^{(2)}_{\alpha\alpha}= c_{\alpha}^2 c_{s\alpha}^2,\\
\end{equation}
\begin{equation}
\mathbf{\Delta}^{(4)}_{\alpha\beta\gamma\theta}=0,else
\end{equation}
\end{subequations}
which can be used to obtain
\begin{subequations}\label{eq:3-3-8}
\begin{equation}
\omega_7=\frac{c_{s1}^2 c_{s2}^2}{4c_1^2 c_2^2}-2\omega_{19},\omega_{11}=\frac{c_{s1}^2 c_{s3}^2}{4c_1^2 c_3^2}-2\omega_{19},\omega_{15}=\frac{c_{s2}^2 c_{s3}^2}{4c_2^2 c_3^2}-2\omega_{19},
\end{equation}
\begin{equation}
\omega_1=\frac{c_{s1}^2}{2c_1^2}-2(\omega_7+\omega_{11})-4\omega_{19},
\end{equation}
\begin{equation}
\omega_2=\frac{c_{s2}^2}{2c_2^2}-2(\omega_7+\omega_{15})-4\omega_{19},
\end{equation}
\begin{equation}
\omega_3=\frac{c_{s3}^2}{2c_3^2}-2(\omega_{11}+\omega_{15})-4\omega_{19}.
\end{equation}
\end{subequations}
From Eqs. (\ref{eq:3-3-8}) and (\ref{eq:3-2}), we can derive
\begin{eqnarray}
& \left\{\begin{array}{c}0<2 \omega_1+4(\omega_7+\omega_{11})+8\omega_{19}=c_{s1}^2/c_{1}^2<1, \\
0<2 \omega_2+4(\omega_7+\omega_{15})+8\omega_{19}=c_{s2}^2/c_{2}^2<1, \\
0<2 \omega_3+4(\omega_{11}+\omega_{15})+8\omega_{19}=c_{s3}^2/c_{3}^2<1.
\end{array}\right.
\end{eqnarray}
 Then we have $0 < \omega_{19}<c_{s\alpha}^2/(8 c_{\alpha}^2)$ $(\alpha=1,2,3)$, while for some other lattice models, such as rD3Q19 lattice, $\omega_{19}$ can be equal to $0$.

From Eq. (\ref{eq:3-3-8}), we can derive some special cases of the rD3Q27 lattice model.

(1) If we take $\omega_{19}=0$, the weight coefficients in the rD3Q19 lattice model are obtained.

(2) If we take $\omega_7=\omega_{11}=\omega_{15}=0$, the weight coefficients in the rD3Q15 lattice model and the following relations can be obtain,
\begin{equation}\label{eq:3-3-9}
\frac{c_{s1}^2}{c_1^2}=\frac{c_{s2}^2}{c_2^2}=\frac{c_{s3}^2}{c_3^2}.
\end{equation}

(3) If we take $\omega_1=\omega_2=\omega_3=\omega_{19}=0$, the weight coefficients in the rD3Q13 lattice model and the following conditions can be derived,
\begin{equation}\label{eq:3-3-10}
\frac{c_{s1}^2}{c_1^2}=\frac{c_{s2}^2}{c_2^2}=\frac{c_{s3}^2}{c_3^2}=\frac{1}{2}.
\end{equation}

It can be found from Eqs. (\ref{eq:3-3-9}) and (\ref{eq:3-3-10}) that when $c_{s1}=c_{s2}=c_{s3}$, one can get $c_1=c_2=c_3$, which implies that there are no rD3Q15 and rD3Q13 lattice models.

From Eqs. (\ref{eq:3-3-4}), (\ref{eq:3-3-6}) and (\ref{eq:3-3-7}), we can give some constraints on the fourth-order moment $\mathbf{\Delta}^{(4)}$ of discrete velocities on the rD$d$Q$q$ lattice,
\begin{equation}\label{eq:3-3-11}
\mathbf{\Delta}^{(4)}_{\alpha\alpha\alpha\alpha}= c_{\alpha}^2 c_{s\alpha}^2,\\ \mathbf{\Delta}^{(4)}_{\alpha\alpha\beta\beta}=\mathbf{\Delta}^{(2)}_{\alpha\alpha}\mathbf{\Delta}^{(2)}_{\beta\beta}=c_{s\alpha}^2 c_{s\beta}^2  \\\ (\alpha\neq \beta),\\  \mathbf{\Delta}^{(4)}_{\alpha\beta\gamma\theta}=0 \\\ (else),
\end{equation}
which indicates that it no longer satisfies the isotropic condition on rectangular lattice or on standard lattice with $c_{s\alpha}^2=c_s^2\neq c^2/3$. Here we present its natural expression as the extension to that on the standard lattice:
\begin{equation}\label{eq:3-7}
\mathbf{\Delta}^{(4)}=<\mathbf{\Delta}^{(2)}\mathbf{\Delta}^{(2)}>+\mathbf{\delta}^{(4)},
\end{equation}
where
\begin{subequations}\label{eq:3-8}
\begin{equation}
<\mathbf{\Delta}^{(2)}\mathbf{\Delta}^{(2)}>_{\alpha\beta\gamma\theta}=\Delta_{\alpha\beta}^{(2)}\Delta_{\gamma\theta}^{(2)}+\Delta_{\alpha\gamma}^{(2)}\Delta_{\beta\theta}^{(2)}+\Delta_{\beta\gamma}^{(2)}\Delta_{\alpha\theta}^{(2)},\\
\end{equation}
\begin{equation}
\mathbf{\delta}^{(4)}_{\alpha\beta\gamma\theta}=c_{s\alpha}^2(c_\alpha^2-3c_{s\alpha}^2),\alpha=\beta=\gamma=\theta;\\
\mathbf{\delta}^{(4)}_{\alpha\beta\gamma\theta}=0, else.
\end{equation}
\end{subequations}

When the relations $c_{s\alpha}=c_s$ and $c_\alpha^2=c^2=3c_{s}^2$ are satisfied for all $\alpha$, one can obtain $\delta^{(4)}=0$ and $\mathbf{\Delta}^{(4)}=c_s^4\mathbf{\Delta}$ with $\Delta_{\alpha\beta\gamma\theta}=\delta_{\alpha\beta}\delta_{\gamma\theta}+\delta_{\alpha\gamma}\delta_{\beta\theta}+\delta_{\beta\gamma}\delta_{\alpha\theta}$, which means that the fourth-order isotropy can be achieved.

Based on above analysis and our previous work  \cite{LuChaiShi2011}, we now present the REDF with general form on a rD$d$Q$q$ lattice, which is an extension of the commonly used one,
\begin{equation}\label{eq:3-3}
g_j^{eq}(A,\mathbf{B},\mathbf{M})=\omega_j\left(A+\tilde{\textbf{c}}_j\cdot{\mathbf{B}}
                   +\tilde{\textbf{Q}}_j:\mathbf{M}\right),
\end{equation}
where
\begin{equation}\label{eq:3-4}
\mathbf{\tilde{c}}_{j\alpha}=\mathbf{c}_{j\alpha}/c_{s\alpha}^2,
\mathbf{\tilde{Q}}_{j\alpha\alpha}=\mathbf{Q}_{j\alpha\alpha}/(c_{s\alpha}^2(c_\alpha^2-c_{s\alpha}^2)),
\mathbf{\tilde{Q}}_{j\alpha\beta}=\mathbf{Q}_{j\alpha\beta}/(2 c_{s\alpha}^2 c_{s\beta}^2) \ \ (\alpha \neq\beta),
\mathbf Q_j=\mathbf c_j \mathbf c_j -\mathbf \Delta^{(2)}.
\end{equation}

According to the properties of rD$d$Q$q$ lattice models, and through some algebra manipulations, we can obtain the basic moments of $g_j^{eq}$ on rD$d$Q$q$ lattice \cite{LuChaiShi2011},
\begin{subequations}\label{eq:3-5}
\begin{equation}
\sum_j g_j^{eq}=A,\ \ \sum_j \mathbf{c}_j g_j^{eq}=\mathbf{B},\ \ \sum_j \mathbf{c}_j \mathbf{c}_j
g_j^{eq}=A \mathbf{\Delta}^{(2)}+\mathbf{M},
\end{equation}
\begin{equation}
\sum_j \mathbf{c}_j \mathbf{c}_j\mathbf{c}_j g_j^{eq}=\mathbf{\Delta}^{(4)}\cdot \tilde{\mathbf{B}},
\end{equation}
\end{subequations}
where $\tilde{\mathbf{B}}_{\alpha}=\mathbf B_{\alpha}/ c_{s\alpha}^2$.

With the help of Eq. (\ref{eq:3-7}), we can rewrite Eq. (\ref{eq:3-5}b) as
\begin{equation}\label{eq:3-9}
\sum_j \mathbf{c}_{j\alpha} \mathbf{c}_{j\beta}\mathbf{c}_{j\gamma} g_j^{eq}=\Delta_{\alpha\beta}^{(2)}B_\gamma+\Delta_{\alpha\gamma}^{(2)}B_\beta+\Delta_{\beta\gamma}^{(2)}B_\alpha+\delta^{(4)}_{\alpha\beta\gamma\theta} \tilde{B}_\theta.
\end{equation}

\textbf{Remark} \uppercase\expandafter{\romannumeral1}: For the rD2Q5I, rD2Q5II, rD3Q7 and rD3Q9 lattice models, $\mathbf{\Delta}^{(4)}_{\alpha\alpha\beta\beta}=c_{s\alpha}^2 c_{s\beta}^2$ ($\alpha\neq \beta$) does not hold.
For the rD2Q5I lattice, $\omega_5=0$, we have $\omega_{\alpha}=\frac{c_{s\alpha}^2}{2c_{\alpha}^2}$ ($\alpha=1,2$), while for the rD2Q5II lattice, $\omega_1=\omega_2=0$, we can obtain $\omega_5=\frac{c_{s\alpha}^2}{4c_{\alpha}^2}$ ($\alpha=1,2$) and also $\frac{c_{s1}^2}{c_1^2}=\frac{c_{s2}^2}{c_2^2}$.
For the rD3Q9 lattice, $\omega_\alpha=0$ ($1\leq\alpha \leq 18$), we have $\omega_{19}=\frac{c_{s\alpha}^2}{4c_{\alpha}^2}$ ($\alpha=1,2,3$) and $\frac{c_{s1}^2}{c_1^2}=\frac{c_{s2}^2}{c_2^2}=\frac{c_{s3}^2}{c_3^2}$.
For the rD3Q7 lattice, $\omega_\alpha=0$ ($\alpha>6$), we derive $\omega_{\alpha}=\frac{c_{s\alpha}^2}{2c_{\alpha}^2}$ ($\alpha=1,2,3$).
Usually, the linear EDF $g_j^{eq}(A,\mathbf{B},\mathbf{0})$ is used, which satisfies Eq. (\ref{eq:3-5}) with $\mathbf{M}=0$, while $\mathbf{\Delta}^{(4)}$ does not satisfy Eq. (\ref{eq:3-3-11}) or Eq. (\ref{eq:3-7}) since $\mathbf{\Delta}^{(4)}_{\alpha\alpha\beta\beta}=c_{s\alpha}^2 c_{s\beta}^2$ ($\alpha\neq \beta$) does not necessarily hold.

\textbf{Remark} \uppercase\expandafter{\romannumeral2}: If we take $\omega_0=0$ or remove the rest velocity $\mathbf{c}_0=\mathbf{0}$ in the rD$d$Q$q$ lattice model mentioned above, a rD$d$Q$(q-1)$ lattice model can be obtained.

\section{The analysis of RMRT-LB method: Direct Taylor expansion}
\label{printlayout}

There are four basic analysis methods for LB models that can be used to recover the macroscopic NSEs and NCDE from the LB models, i.e., the Chapman-Enskog (CE) analysis, the Maxwell iteration (MI) method, the direct Taylor expansion (DTE) method and the recurrence equations (RE) method. In Ref.  \cite{Chai2020}, these four methods have been compared, and it is shown that they can give the same equations at the second-order of expansion parameters, while the DTE method is much simpler. In what follows, only the DTE method is used to analyze the RMRT-LB model for NSEs.

Applying the Taylor expansion to Eq. (\ref{eq:2-1}),  one can get
\begin{equation}\label{eq:4-1}
\sum_{l=1}^{N}\frac{\Delta t^l}{l!}D_j^l f_j +O(\Delta t^{N+1})=-\mathbf{\Lambda}_{jk} f_k^{ne}+\Delta t \tilde{F}_j, N\geq 1,
\end{equation}
where $\tilde{F}_j=G_j+F_j+\Delta t \bar{D}_j F_j/2$.

Based on $f_j=f_j^{eq}+f_j^{ne}$ and Eq. (\ref{eq:4-1}), the following equations can be obtained,
\begin{subequations}\label{eq:4-2}
\begin{equation}\label{eq:4-2a}
f_j^{ne}=O(\Delta t),
\end{equation}
\begin{equation}\label{eq:4-2b}
\sum_{l=1}^{N-1}\frac{\Delta t^l}{l!}D_j^l (f_j^{eq}+f_j^{ne}) + \frac{\Delta t^N}{N!}D_j^N f_j^{eq}=-\mathbf{\Lambda}_{jk} f_k^{ne}+\Delta t \tilde{F}_j+O(\Delta t^{N+1}), N\geq 1.
\end{equation}
\end{subequations}
Then from Eq. (\ref{eq:4-2b}), we can derive the equations at different orders of $\Delta t$,

\begin{subequations}\label{eq:4-3}
\begin{equation}\label{eq:4-3a}
D_j f_j^{eq}=-\frac{\mathbf{\Lambda}_{jk}}{\Delta t } f_k^{ne}+G_j+F_j+O(\Delta t),
\end{equation}
\begin{equation}\label{eq:4-3b}
D_j (f_j^{eq}+f_j^{ne})+\frac{\Delta t}{2}D_j^2 f_j^{eq}=-\frac{\mathbf{\Lambda}_{jk}}{\Delta t } f_k^{ne}+ G_j+F_j+\frac{\Delta t}{2} \bar{D}_jF_j+O(\Delta t^2).
\end{equation}
\end{subequations}
According to Eq. (\ref{eq:4-3a}), we have
\begin{equation}\label{eq:4-4}
\frac{\Delta t}{2}D_j^2 f_j^{eq}=-\frac{1}{2} D_j \mathbf{\Lambda}_{jk} f_k^{ne}+\frac{\Delta t}{2}D_j(G_j+F_j)+O(\Delta t^2),
\end{equation}
Substituting Eq. (\ref{eq:4-4}) into Eq. (\ref{eq:4-3b}), one can obtain the following equation,
\begin{equation}\label{eq:4-5}
D_j f_j^{eq}+D_j\big(\delta_{jk}-\frac{\mathbf{\Lambda}_{jk}}{2}\big)f_k^{ne}+\frac{\Delta t}{2}D_j G_j =-\frac{\mathbf{\Lambda}_{jk}}{\Delta t} f_k^{ne}+G_j+F_j+\frac{\Delta t}{2} (\bar{D}_j-D_j)F_j+O(\Delta t^2).
\end{equation}

Based on Eqs. (\ref{eq:4-3a}) and (\ref{eq:4-5}) with proper constraints on the collision matrix $\mathbf{\Lambda}$ and the moments of $f_j^{eq}$, $G_j$ and $F_j$, the related macroscopic equation (NSEs and NCDE) can be recovered. For NSEs, if we take $\gamma=1$, then Eq. (\ref{eq:4-5}) can be simplified to
\begin{equation}\label{eq:4-6}
D_j f_j^{eq}+D_j\big(\delta_{jk}-\frac{\mathbf{\Lambda}_{jk}}{2}\big)f_k^{ne}+\frac{\Delta t}{2}D_j G_j =-\frac{\mathbf{\Lambda}_{jk}}{\Delta t} f_k^{ne}+G_j+F_j+O(\Delta t^2).
\end{equation}

For the sake of convenience, we introduce the following matrices,
\begin{subequations}\label{eq:4-7}
\begin{equation}
\mathbf{e}=(1,1,\cdots,1)=(\mathbf{e}_k)_{1\times q},
\end{equation}
\begin{equation}
\mathbf{E}=(\mathbf{c}_0,\mathbf{c}_1,\cdots,\mathbf{c}_{q-1})=(\mathbf{c}_k)_{d\times q},
\end{equation}
\begin{equation}
\langle \mathbf{EE} \rangle=(\mathbf{c}_0\mathbf{c}_0,\mathbf{c}_1\mathbf{c}_1,\cdots,\mathbf{c}_{q-1}\mathbf{c}_{q-1})=(\mathbf{c}_k\mathbf{c}_k)_{d^2\times q},
\end{equation}
\end{subequations}
where $\mathbf{e}_k$, $\mathbf{c}_k$ and $\mathbf{c}_k\mathbf{c}_k$ are the $k$-th column of $\mathbf{e}$, $\mathbf{E}$ and $\langle \mathbf{EE} \rangle$, respectively.

\subsection{The nonlinear convection-diffusion equation}

The $d$-dimensional NCDE with a source term considered in this work can be expressed as  \cite{Chai2016a,Chai2020}
\begin{equation}\label{eq:NCDE}
\partial_t \phi+\nabla\cdot\mathbf{B}= \nabla\cdot[\mathbf{K}\cdot(\nabla\cdot \mathbf{D})]+S,
\end{equation}
where $\phi$ is a unknown scalar function of position $\mathbf{x}$ and time $t$, $S$ is a scalar source term. $\mathbf{B}=(\mathbf{B}_\alpha)$ is a vector function, $\mathbf{K}=(\mathbf{K}_{\alpha\beta})$ and $\mathbf{D}=(\mathbf{D}_{\alpha\beta})$ are symmetric tensors (matrices), and they can be functions of $\phi, \mathbf{x}$, and $t$. It should be noted that Eq. (\ref{eq:NCDE}) can be considered as a general form of CDEs, and many different kinds of CDEs considered in some previous work  \cite{LuChaiShi2011,Ginzburg2005a,Shi2009,Chai2019} are its special cases.

To recover Eq. (\ref{eq:NCDE}) from the RMRT-LB method (\ref{eq:2-1}), the unknown function $\phi$ can be calculated by $\phi=\sum_j f_j$ with the mass conservation $\sum_j f_j^{ne}= \sum_j (f_j-f_j^{eq})=0$, and the collision matrix $\mathbf{\Lambda}$ and moments of $f_j^{eq}$, $G_j$ and $F_j$ should satisfy the following relations  \cite{Chai2020},
\begin{subequations}\label{eq:4-9}
\begin{equation}
\sum_j \mathbf{e}_{j}\mathbf{\Lambda}_{jk} = s_0 \mathbf{e}_{k},\ \ \sum_j \mathbf{c}_{j}\mathbf{\Lambda}_{jk} = \mathbf{S}_{10}\mathbf{e}_{k}+ \mathbf{S}_1 \mathbf{c}_k,\ \ \forall k,
\end{equation}
\begin{equation}
\sum_j f_j^{eq}=\phi,\ \ \sum_j \mathbf{c}_j f_j^{eq}=\mathbf{B},\ \ \sum_j \mathbf{c}_j \mathbf{c}_j
f_j^{eq}= \beta c_s^2\mathbf{D}+\mathbf{C},
\end{equation}
\begin{equation}
\sum_j F_j=S,\ \ \sum_j \mathbf{c}_j F_j=\mathbf{0},\ \ \
\sum_j G_j=0,\ \ \sum_j \mathbf{c}_j G_j=\mathbf{M}_{1G},
\end{equation}
\end{subequations}
where $\mathbf{S}_{10}$ is a $d\times1$ matrix, $\mathbf{S}_1$ is an invertible $d\times d$ relaxation matrix corresponding to the diffusion matrix $\mathbf{K}$, $\beta$ is a parameter for adjusting the relaxation matrix $\mathbf{S}_1$ in Eq. (\ref{eq:4-10}), $\mathbf{C}$ is an auxiliary moment, and $\mathbf{M}_{1G}$ is the first-order moment of $G_j$.

Note that Eq. (\ref{eq:4-9}) is the same as that of the standard MRT-LB model for NCDE (\ref{eq:NCDE}) in Ref. \cite{Chai2020}, here we omit the same process in recovering the macroscopic NCDE through the DTE analysis, and only give some useful formulas below, including the diffusion tensor $\mathbf{K}$, $\mathbf{M}_{1G}$ and the diffusion flux $-\mathbf{K}\cdot(\nabla\cdot \mathbf{D})$,
\begin{equation}\label{eq:4-10}
\mathbf{K}=\Delta t \beta c_s^2(\mathbf{S}_1^{-1}-\mathbf{I}/2),
\end{equation}
\begin{eqnarray}\label{eq:4-11}
\mathbf{M}_{1G} & = & (\mathbf{I}-\mathbf{S}_1/2)(\partial_t\mathbf{B}+\nabla\cdot \mathbf{C})\nonumber\\
& = & \left\{\begin{array}{c} (\mathbf{I}-\mathbf{S}_1/2)\mathbf{B}'S,\ \ \ if\ \mathbf{B}=\mathbf{B(\phi)}\ and \ \mathbf{C}=\int\mathbf{B}'\mathbf{B}'d\phi,\\
 (\mathbf{I}-\mathbf{S}_1/2)\partial_{t}\mathbf{B},\ \ \ \ \ \ \ \ \ \ \ \ \ \  \ \ \ \ \ \ \ \ \ \ \ \ \ \ \ \ \ \ \ \ \ \ \ \ \  \ if \ \mathbf{C}=\mathbf{0},
\end{array}\right.
\end{eqnarray}
\begin{eqnarray}\label{eq:4-13}
-\mathbf{K}\cdot(\nabla\cdot\mathbf{D}) = & \left\{\begin{array}{c} (\mathbf{I}-\mathbf{S}_1/2)\big(\mathbf{Ef}^{ne}+\Delta t\mathbf{B}'S/2\big),\ \ if\ \mathbf{B}=\mathbf{B(\phi)}\ and \ \mathbf{C}=\int\mathbf{B}'\mathbf{B}'d\phi, \\
(\mathbf{I}-\mathbf{S}_1/2)\big(\mathbf{Ef}^{ne}+\Delta t\partial_t \mathbf{B}/2\big),\ \ \ \ \ \ \ \ \ \ \ \ \ \  \ \ \ \ \ \ \ \ \ \ \ \ \ \ \ \ \ \ \ \ \ \ \ \  \ if\ \mathbf{C}=\mathbf{0},
\end{array}\right.
\end{eqnarray}
where $\mathbf{B}'=\frac{d\mathbf{B}}{d\phi}$, $\mathbf{f}^{ne}=(f_0^{ne},f_1^{ne},\cdots,f_{q-1}^{ne})^{T}$. Additionally, if $\mathbf{D}$ is only a function of $\phi$, we can also obtain a local scheme for $\nabla\phi$ from Eq. (\ref{eq:4-13}).

Actually, Eq. (\ref{eq:4-13}) can be written as
\begin{eqnarray}\label{eq:4-13-1}
\mathbf{Ef}^{ne} = & \left\{\begin{array}{c} -(\mathbf{I}-\mathbf{S}_1/2)^{-1}\big(\mathbf{K}\cdot(\nabla\cdot\mathbf{D})\big)-  \Delta t\mathbf{B}'S/2,\ \ if\ \mathbf{B}=\mathbf{B(\phi)}\ and \ \mathbf{C}=\int\mathbf{B}'\mathbf{B}'d\phi, \\
-(\mathbf{I}-\mathbf{S}_1/2)^{-1}\big( \mathbf{K}\cdot(\nabla\cdot\mathbf{D}) \big)-\Delta t\partial_t \mathbf{B}/2,\ \ \ \ \ \ \ \ \ \ \ \ \ \  \ \ \ \ \ \ \ \ \ \ \ \ \ \ \ \ \ \ \ \ \ \ \ \  \ if\ \mathbf{C}=\mathbf{0}.
\end{array}\right.
\end{eqnarray}
In addition, if we denote $\mathbf{m}_1^{ne}=\mathbf{Ef}^{ne}$, and consider Eq. (\ref{eq:4-13-1}) and the mass conservation $\mathbf{ef}^{ne}=\sum_j f_j^{ne}=0$, one can obtain a useful formula to approximate $f_j^{ne}$,
\begin{equation}\label{eq:4-13-2}
f_j^{ne}=g_j^{eq}(0,\mathbf{m}_1^{ne},\mathbf{0})=\omega_j \frac{\mathbf{c}_j\cdot \mathbf{m}_1^{ne}}{c_s^2}=\omega_j \frac{\mathbf{c}_{j\alpha} \mathbf{m}_{1,\alpha}^{ne}}{c_s^2},
\end{equation}
which can be used to initialize the distribution function $f_j$.

\subsection{The equilibrium, auxiliary and source distribution functions of RMRT-LB method for NCDE}

As discussed above, it can be found that to recover the NCDE (\ref{eq:NCDE}) correctly, some proper requirements on the equilibrium, auxiliary and source distribution functions should be satisfied. Based on Eqs. (\ref{eq:3-3}), (\ref{eq:3-4}) and (\ref{eq:4-9}) for a rD$d$Q$q$ lattice model, we can obtain the following expressions of $f_{j}^{eq}$, $G_j$ and $F_j$ with $c_{s \alpha}=c_s$ for all $\alpha$,
\begin{eqnarray}\label{eq:4-14}
f_j^{eq} & = & g_j^{eq}(\phi, \mathbf{B}, \beta c_s^2\mathbf{D}+\mathbf{C}- c_s^2 \phi\mathbf{I}) \nonumber\\
         & = & \omega_j\left[\phi+\frac{\textbf{c}_{j\alpha}{\mathbf{B}_{\alpha}}}{c_s^2}
                   +\frac{(\beta c_s^2\mathbf{D}+\mathbf{C}- c_s^2 \phi\mathbf{I})_{\alpha\alpha}(c_{j\alpha}^2 - c_s^2)}{c_s^2(c_{\alpha}^2-c_s^2)}+\frac{(\beta c_s^2\mathbf{D}+\mathbf{C}-  c_s^2 \phi\mathbf{I})_{\alpha\bar{\alpha}}(c_{j\alpha}c_{j\bar{\alpha}})}{2c_s^4}\right],
\end{eqnarray}
\begin{equation}\label{eq:4-14-1}
G_j=g_j^{eq}(0,\mathbf{M}_{1G},\mathbf{0})=\omega_j \frac{\mathbf{c}_{j\alpha} \mathbf{M}_{1G,\alpha}}{c_s^2},\\\ F_j=g_j^{eq}(S,\mathbf{0},\mathbf{0})=\omega_j S,
\end{equation}
where $\bar{\alpha}$ denotes the index $\gamma$ such that $\gamma\neq\alpha$, $\mathbf{M}_{1G}$ is given by Eq. (\ref{eq:4-11}).

We note that if $\mathbf{D}=\phi \mathbf{I}$, the NCDE (\ref{eq:NCDE}) would become
\begin{equation}\label{eq:4-15}
\partial_t \phi+\nabla\cdot\mathbf{B}= \nabla\cdot(\mathbf{K}\cdot\nabla\phi)+S.
\end{equation}
If we further take $\beta =1$ and $\mathbf{C}=0$, $f_j^{eq}$, $G_j$ and $F_j$ defined by Eqs. (\ref{eq:4-14}) and (\ref{eq:4-14-1}) can be simplified as
\begin{subequations}\label{eq:4-16}
\begin{equation}
f_j^{eq}=g_j^{eq}(\phi,\mathbf{B},\mathbf{0})=\omega_j \left[\phi+\frac{\mathbf{c}_{j\alpha} \mathbf{B}_{\alpha}}{c_s^2}\right],
\end{equation}
\begin{equation}
G_j=g_j^{eq}(\mathbf{0},(\mathbf{I}-\mathbf{S}_1/2)\partial_t \mathbf{B},\mathbf{0})=\omega_j \left[\frac{\mathbf{c}_{j\alpha} ((\mathbf{I}-\mathbf{S}_1/2)\partial_t \mathbf{B})_{\alpha}}{c_s^2}\right],
F_j=g_j^{eq}(S,\mathbf{0},\mathbf{0})=\omega_j S,
\end{equation}
\end{subequations}
where the term $\partial_t \mathbf{B}$ can be computed by the first-order explicit difference scheme, i.e., $\partial_t \mathbf{B}=[\mathbf{B}(\mathbf{x}, t)-\mathbf{B}(\mathbf{x}, t-\Delta t)]/\Delta t$. In this case, the simpler rD$d$Q$2d$ or rD$d$Q($2d+1$) lattice model can be adopted.

In addition, we would also like to point out that at the diffusive scaling, the present RMRT-LB model can be simplified \cite{Chai2020}, i.e., the term $\frac{\Delta t^2}{2}\bar{D}_j F_j(\mathbf{x}, t)$ in the evolution equation [Eq. (\ref{eq:2-1})] can be removed, and Eqs. (\ref{eq:4-14}) and (\ref{eq:4-14-1}) would reduce to
\begin{equation}\label{eq:4-17}
f_j^{eq}=g_j^{eq}(\phi, \mathbf{B}, \beta c_s^2\mathbf{D}-c_s^2\phi \mathbf{I}),\\\ G_j=0,\\\ F_j=\omega_j S.
\end{equation}

\textbf{Remark} \uppercase\expandafter{\romannumeral1}: The first-order moment of $f_j^{ne}$, i.e. $\mathbf{Ef}^{ne}$ in Eq. (\ref{eq:4-13}) can also be computed by its counterpart ($\mathbf{Ef}^{ne}=\mathbf{m}_\mathbf{E}^{ne}=\mathbf{m}_\mathbf{E}-\mathbf{m}_\mathbf{E}^{eq}$) in the moment space.

\textbf{Remark} \uppercase\expandafter{\romannumeral2}: When $\mathbf{K}=\mathbf{diag}(\kappa_{\alpha})$ and $\mathbf{D}=\mathbf{diag}( D_{\alpha})$ are diagonal matrices, the term $\beta c_s^2 \mathbf{D}$ in Eq. (\ref{eq:4-9}b) can also be replaced by $\mathbf{diag}(\beta_{\alpha}c_{s\alpha}^2 D_{\alpha})$, and Eq. (\ref{eq:4-10}) becomes
\begin{equation}\label{eq:4-18-0}
\kappa_{\alpha}=\Delta t (\tau-1/2)\beta_{\alpha}c_{s\alpha}^2,
\end{equation}
where $\mathbf{S}_1^{-1}=\tau \mathbf{I}$ has been used. In this case, the SRT-RLB model can be used for NCDE (\ref{eq:NCDE}), as shown in Ref. \cite{LuChaiShi2011}.

\textbf{Remark} \uppercase\expandafter{\romannumeral3}: If $c_s^2\neq c^2/3 $ is considered on the standard lattice, the distribution functions $f_j^{eq}, G_j$ and $F_j$ defined in Eqs. (\ref{eq:4-14}), and (\ref{eq:4-14-1}) for NCDE are new, and one get a class of
MRT-LB models as the special cases of the present RMRT-LB model with the sound speed $c_s^2$ as an adjustable parameter .

\subsection{The Navier-Stokes equations}

We now consider the following $d$-dimensional NSEs with the source and forcing terms,
\begin{subequations}\label{eq:NSE}
\begin{equation}
\partial_t \rho+\nabla\cdot(\rho \mathbf{u})= \bar{S},
\end{equation}
\begin{equation}
\partial_t (\rho\mathbf{u})+\nabla\cdot(\rho \mathbf{uu})=-\nabla p+ \nabla\cdot \mathbf{\sigma}+ \bar{\mathbf{F}},
\end{equation}
\end{subequations}
where the shear stress $\mathbf{\sigma}$ is defined by
\begin{equation}
\mathbf{\sigma}=\mu\big[\nabla \mathbf{u}+(\nabla \mathbf{u})^T\big]+\lambda (\nabla\cdot \mathbf{u})\mathbf{I}=\mu\big[\nabla \mathbf{u}+(\nabla \mathbf{u})^T-\frac{2}{d}(\nabla\cdot \mathbf{u})\mathbf{I}\big]+\mu_{b} (\nabla\cdot \mathbf{u})\mathbf{I},
\end{equation}
where $\mu$ is the dynamic viscosity, $\lambda=\mu_{b}-2\mu/d$ with $\mu_{b}$ being the bulk viscosity  \cite{Dellar2001,Kundu2016}.

Similar to the derivation of NCDE, to recover NSEs from MRT-LB method (\ref{eq:2-1}) we also need to give some constraints on $\mathbf{\Lambda}$, $f_j$, $f_j^{eq}$, $G_j$, and $F_j$. In addition, compared to Eq. (\ref{eq:4-9}), there is another requirement on the high-order moments of the distribution function for NSEs. Here the following conditions should be satisfied,
\begin{subequations}\label{eq:4-18}
\begin{equation}
\rho=\sum_j f_j=\sum_j f_j^{eq},\ \ \rho \mathbf{u}=\sum_j \mathbf{c}_j f_j=\sum_j \mathbf{c}_j f_j^{eq},
\end{equation}
\begin{equation}
\sum_j \mathbf{c}_j \mathbf{c}_j f_j^{eq}= \rho \mathbf{\Delta}^{(2)}+\rho\mathbf{uu},\ \
\sum_j \mathbf{c}_j \mathbf{c}_j \mathbf{c}_j f_j^{eq}= \rho \mathbf{\Delta}^{(4)} \cdot \mathbf{\tilde{u}},
\end{equation}
\begin{equation}
\sum_j F_j=\bar{S},\ \ \sum_j \mathbf{c}_j F_j=\bar{\mathbf{F}},\ \ \sum_j \mathbf{c}_j \mathbf{c}_j F_j=\mathbf{0},
\end{equation}
\begin{equation}
\sum_j G_j=0,\ \ \sum_j \mathbf{c}_j G_j=\mathbf{0},\ \ \sum_j \mathbf{c}_j \mathbf{c}_j G_j=\mathbf{M}_{2G},
\end{equation}
\begin{equation}
\sum_j\mathbf{e}_j\mathbf{\Lambda}_{jk} = s_0\mathbf{e}_k, \ \  \sum_j \mathbf{c}_j\mathbf{\Lambda}_{jk} = \mathbf{S}_{10} \mathbf{e}_k + \mathbf{S}_1 \mathbf{c}_k,\ \ \
\sum_j\mathbf{c}_j\mathbf{c}_j\mathbf{\Lambda}_{jk} = \sum_j\mathbf{S}_{20} \mathbf{e}_k + \mathbf{S}_{21} \mathbf{c}_k+\mathbf{S}_{2}\mathbf{c}_k \mathbf{c}_k,\ \ \
\end{equation}
\end{subequations}
where $\mathbf{M}_{2G}$ is a second-order tensor to be determined below, $\mathbf{\Delta}^{(4)}=<\mathbf{\Delta}^{(2)}\mathbf{\Delta}^{(2)}>+\delta^{(4)}$ is defined in Eq. (\ref{eq:3-7}), $\mathbf{\tilde{u}}_{\alpha}=\mathbf{u}_{\alpha}/c_{s\alpha}^2$, $\mathbf{S}_{10}$ is a $d\times 1$ matrix,
$\mathbf{S}_1$ is an invertible $d\times d$ relaxation sub-matrix, $\mathbf{S}_{20}$ and $\mathbf{S}_{21}$ are two $d^{2}\times1$ and $d^{2}\times d$ matrices, and $\mathbf{S}_2$ is an invertible $d^2\times d^2$ relaxation sub-matrix corresponding to the dynamic and bulk viscosities. Additionally, Eq. (\ref{eq:4-18}a) gives the following conditions,
\begin{equation}\label{eq:4-18-1}
\sum_j f_j^{ne}=\sum_j (f_j-f_j^{eq})=0,\ \ \  \sum_j \mathbf{c}_j f_j^{ne}=\sum_j \mathbf{c}_j (f_j-f_j^{eq})=\mathbf{0},
\end{equation}
which means the mass and momentum conservation.

Now if we take $c_{s\alpha}^2=c_s^2$ for all $\alpha$, then Eq. (\ref{eq:4-18}b) becomes
\begin{equation}\label{eq:4-18-2}
\sum_j \mathbf{c}_j \mathbf{c}_j f_j^{eq}= c_s^2\rho \mathbf{I}+\rho\mathbf{uu},\ \
\sum_j \mathbf{c}_j \mathbf{c}_j \mathbf{c}_j f_j^{eq}= \rho(c_s^2 \mathbf{\mathbf{\Delta}}+\bar{\delta}^{(4)}) \cdot \mathbf{u},
\end{equation}
where $\mathbf{\bar{\delta}}^{(4)}=\mathbf{\delta}^{(4)}/c_s^2$.

Summing Eq. (\ref{eq:4-3}a) and adopting Eqs. (\ref{eq:4-18}) and (\ref{eq:4-18-1}), we can obtain
\begin{equation}\label{eq:4-19}
\partial_t \rho+\nabla\cdot (\rho\mathbf{u}) = \bar{S}+O(\Delta t^2),
\end{equation}
which indicates that the continuity equation (\ref{eq:NSE}a) is recovered correctly at the order of $O(\Delta t^2)$.

Multiplying $\mathbf{c}_j$ on both sides of Eqs. (\ref{eq:4-3}a) and (\ref{eq:4-6}), and through a summation over $j$, we have
\begin{subequations}\label{eq:4-20}
\begin{equation}
\partial_t (\rho\mathbf{u})+\nabla\cdot ( c_s^2 \rho \mathbf{I}+\rho\mathbf{uu}) = \bar{\mathbf{F}}+O(\Delta t),
\end{equation}
\begin{eqnarray}
\partial_t (\rho\mathbf{u})+\nabla\cdot (c_s^2 \rho \mathbf{I}+\rho\mathbf{u}\mathbf{u})  -  \nabla\cdot (\mathbf{I}-\mathbf{S}_2 /2)\sum_k\mathbf{c}_k\mathbf{c}_k f_k^{(ne)} + \frac{\Delta t}{2}\nabla\cdot \mathbf{M}_{2G}=  \bar{\mathbf{F}}+O(\Delta t^2),
\end{eqnarray}
\end{subequations}
where Eqs. (\ref{eq:4-18}), (\ref{eq:4-18-1}) and (\ref{eq:4-18-2}) have been used.

Additionally, from Eqs. (\ref{eq:4-3}a), (\ref{eq:4-18}), (\ref{eq:4-18-2}) and (\ref{eq:4-19}) we get
\begin{eqnarray}\label{eq:4-21}
\sum_k \mathbf{c}_k\mathbf{c}_k f_k^{ne} & = &-\Delta t \mathbf{S}_2^{-1}\sum_k \mathbf{c}_k\mathbf{c}_k (D_k f_k^{eq}-G_k-F_k)+O(\Delta t^2)\nonumber\\
& = &-\Delta t\mathbf{S}_2^{-1}\big[\partial_t \sum_k \mathbf{c}_k\mathbf{c}_k f_k^{eq}+\nabla\cdot \sum_k \mathbf{c}_k\mathbf{c}_k\mathbf{c}_k f_k^{eq}-\mathbf{M}_{2G}\big]+O(\Delta t^2)\nonumber\\
& = &-\Delta t\mathbf{S}_2^{-1}\big[\partial_t (c_s^2\rho \mathbf{I}+\rho\mathbf{uu})+\nabla\cdot (\rho( c_s^2\mathbf{\Delta}+ \bar{\delta}^{(4)})\cdot \mathbf{u})-\mathbf{M}_{2G}\big]+O(\Delta t^2)\nonumber\\
& = &-\Delta t\mathbf{S}_2^{-1}\big[\partial_t (\rho\mathbf{uu})+c_s^2(\nabla \rho \mathbf{u}+(\nabla \rho \mathbf{u})^{T})+\nabla\cdot(\rho\bar{\delta}^{(4)}\cdot\mathbf{u})+c_s^2\bar{S}\mathbf{I}-\mathbf{M}_{2G}\big]+O(\Delta t^2).
\end{eqnarray}
Based on Eqs. (\ref{eq:4-19}) and (\ref{eq:4-20}a), the following equations can be obtained,
\begin{subequations}\label{eq:4-22}
\begin{equation}
\partial_t (\rho\mathbf{uu})=\mathbf{u}\bar{\mathbf{F}}+\bar{\mathbf{F}}\mathbf{u}-c_s^2\big[\mathbf{u}\nabla\rho+(\mathbf{u}\nabla\rho)^T\big]
-\nabla\cdot(\rho\mathbf{uuu})-\mathbf{uu}\bar{S}+O(\Delta t Ma),
\end{equation}
\begin{equation}
c_s^2(\nabla \rho \mathbf{u}+(\nabla \rho \mathbf{u})^T)
=c_s^2 \rho\big[\nabla \mathbf{u}+(\nabla \mathbf{u})^T\big]+c_s^2\big[\mathbf{u}\nabla\rho+(\mathbf{u}\nabla\rho)^T\big],
\end{equation}
\begin{equation}
\nabla\cdot(\rho\bar{\delta}^{(4)}\cdot\mathbf{u})=\rho\nabla\cdot(\bar{\delta}^{(4)}\cdot\mathbf{u})+O(Ma^3),
\end{equation}
\end{subequations}
then we can rewrite Eq. (\ref{eq:4-21}) as
\begin{eqnarray}\label{eq:4-23}
\sum_k \mathbf{c}_{k}\mathbf{c}_{k} f_k^{ne} & = & - \Delta t \mathbf{S}_2^{-1} \big[\rho c_{s}^{2}(\nabla\mathbf{u}+(\nabla\mathbf{u})^{T})+\rho\nabla\cdot (\bar{\delta}^{(4)}\cdot \mathbf{u})+\mathbf{u}\bar{\mathbf{F}}+\bar{\mathbf{F}}\mathbf{u}+(c_s^2\mathbf{I}-\mathbf{uu})\bar{S}
-\mathbf{M}_{2G}\big]\nonumber\\
& + & O(\Delta t^2+\Delta t Ma^{3}),
\end{eqnarray}
where $Ma$ is the Mach number, and the relations $\nabla \rho=O(Ma^2)$, $\mathbf{u}=O(Ma)$, and $\nabla\cdot(\rho\mathbf{uuu})=O(Ma^{3})$ are used.

Substituting Eq. (\ref{eq:4-23}) into Eq. (\ref{eq:4-20}b) and using Eqs. (\ref{eq:4-18}) and (\ref{eq:4-18-2}), we can obtain
\begin{eqnarray}\label{eq:4-24}
& & \partial_t (\rho\mathbf{u}) + \nabla\cdot (c_s^2\rho \mathbf{I} + \rho\mathbf{uu})\nonumber\\
& = &\Delta t\nabla\cdot \big[\rho \big(\mathbf{S}_2^{-1}-\mathbf{I}/2\big) \big(c_s^2\big(\nabla\mathbf{u}+(\nabla\mathbf{u})^{T}\big)+\nabla\cdot(\bar{\delta}^{(4)}\cdot \mathbf{u})\big)\big]+ \bar{\mathbf{F}} + \Delta t \nabla\cdot \mathbf{RH}_2+O(\Delta t^2+\Delta t Ma^{3}),
\end{eqnarray}
where
\begin{equation}\label{eq:4-25}
\mathbf{RH}_2= \big(\mathbf{S}_2^{-1}-\mathbf{I}/2\big)(\mathbf{u}\bar{\mathbf{F}}+\bar{\mathbf{F}}\mathbf{u}+(c_{s}^{2}\mathbf{I}-\mathbf{u}\mathbf{u})\bar{S})-\mathbf{S}_2^{-1}\mathbf{M}_{2G}.
\end{equation}

To obtain the correct NSEs, the relation $\mathbf{RH}_2=0$ should be satisfied, which gives the following equation,
\begin{equation}\label{eq:4-26}
\mathbf{M}_{2G}  = \big(\mathbf{I}-\mathbf{S}_2 /2\big)(\mathbf{u}\bar{\mathbf{F}}+\bar{\mathbf{F}}\mathbf{u}+(c_{s}^{2}\mathbf{I}-\mathbf{u}\mathbf{u})\bar{S}).
\end{equation}
Then Eq. (\ref{eq:4-24}) can be rewrite as
\begin{equation}\label{eq:4-24-1}
\partial_t (\rho\mathbf{u}) + \nabla\cdot (c_s^2\rho \mathbf{I} + \rho\mathbf{uu}) = \nabla\cdot \mathbf{\tau}+ \bar{\mathbf{F}} +O(\Delta t^2+\Delta t Ma^{3}),
\end{equation}
where
\begin{equation}\label{eq:4-24-2}
\mathbf{\tau}=\Delta t\rho \big(\mathbf{S}_2^{-1}-\mathbf{I}/2\big) \big(c_s^2\big(\nabla\mathbf{u}+(\nabla\mathbf{u})^{T}\big)+\nabla\cdot(\bar{\delta}^{(4)}\cdot \mathbf{u})\big),
\end{equation}
which needs to be determined as viscosity term through selecting proper relaxation matrix $\mathbf{S}_2^{-1}$.

Let
\begin{equation}\label{eq:4-27}
\mathbf{S}_2^{-1}=
\left(
\begin{array}{cc}
    \mathbf{S}_2^{(1)} & 0  \\
    0 &  \mathbf{S}_2^{(2)}  \\
\end{array}
\right),
\end{equation}
with
\begin{equation}\label{eq:4-27-1}
\mathbf{S}_2^{(1)}=\mathbf{diag} (s_{s\alpha}^{-1})+\mathbf{a}\mathbf{b}^{T}/d, \mathbf{S}_2^{(2)}=\mathbf{diag}(s_{\alpha\beta}^{-1})_{\alpha \neq \beta},
\end{equation}
where $\mathbf{a}=(a_{\alpha}), \mathbf{b}=(b_{\beta})$ with $a_{\alpha}=(s_{b\alpha }^{-1}-s_{s\alpha}^{-1})(c_{\alpha}^2-c_{s}^2)$ and $b_{\beta}=1/(c_{\beta}^2-c_{s}^2)$, then substituting Eq. (\ref{eq:4-27}) into Eq. (\ref{eq:4-24-2}), one can obtain the generalized shear stress,
\begin{equation}\label{eq:4-28}
\mathbf{\tau}=\mathbf{K}_{\mu} \big[\nabla\mathbf{u}+(\nabla\mathbf{u})^{T}-\frac{2}{d}(\nabla\cdot\mathbf{u})\mathbf{I}\big]+\mathbf{K}_{\mu_b}\big[(\nabla\cdot\mathbf{u})\mathbf{I}\big],
\end{equation}
where $\mathbf{K}_{\mu}$ and $\mathbf{K}_{\mu_b}$ are dynamic and bulk viscosity tensors defined as
\begin{subequations}\label{eq:4-29}
\begin{equation}
\mathbf{K}_{\mu,\alpha\beta}= \big(s_{\alpha \beta}^{-1}-\frac{1}{2}\big) \rho c_s^2\Delta t,\alpha\neq \beta, \ \  \mathbf{K}_{\mu,\alpha\alpha}= \frac{1}{2}\big(s_{s\alpha}^{-1}-\frac{1}{2}\big) \rho(c_\alpha^2- c_s^2)\Delta t,
\end{equation}
\begin{equation}
\mathbf{K}_{\mu_b,\alpha\alpha}= \frac{1}{d}\big(s_{b\alpha }^{-1}-\frac{1}{2}\big) \rho(c_\alpha^2- c_s^2)\Delta t.
\end{equation}
\end{subequations}
When $p=\rho c_s^2$ and the truncation error $O(\Delta t^2+\Delta t Ma^{3})$ is neglected, Eq. (\ref{eq:4-24-1}) becomes
\begin{eqnarray}\label{eq:4-24-3}
\partial_t (\rho\mathbf{u})+\nabla\cdot (p \mathbf{I}+\rho\mathbf{uu}) & = &\nabla\cdot \mathbf{K}_{\mu}\big[\nabla\mathbf{u}+(\nabla\mathbf{u})^{T}-\frac{2}{d}(\nabla\cdot\mathbf{u})\mathbf{I}\big]+\nabla\cdot\mathbf{K}_{\mu_b}\big[(\nabla\cdot\mathbf{u})\mathbf{I}\big]+ \bar{\mathbf{F}},
\end{eqnarray}
which is a generalized NSEs with dynamic and bulk viscosity in the form of tensors defined by Eq. (\ref{eq:4-29}).

If we take $\mathbf{K}_{\mu}=\mu \mathbf{I} $ and $\mathbf{K}_{\mu_b}=\mu_b \mathbf{I}$, then Eq. (\ref{eq:4-24-3}) can be written in the following form,
\begin{eqnarray}\label{eq:4-30}
\partial_t (\rho\mathbf{u})+\nabla\cdot (p \mathbf{I}+\rho\mathbf{uu}) & = &\nabla\cdot\mathbf{\mu}\big[\nabla\mathbf{u}+(\nabla\mathbf{u})^{T}-\frac{2}{d}(\nabla\cdot\mathbf{u})\mathbf{I}\big]+\nabla\cdot\big[\mu_{b}(\nabla\cdot\mathbf{u})\mathbf{I}\big]+ \bar{\mathbf{F}},
\end{eqnarray}
which would reduce to the momentum equation (\ref{eq:NSE}b) with the following dynamic and bulk viscosities,
\begin{subequations}\label{eq:4-31}
\begin{equation}
\mu= \big(s_{\alpha \beta}^{-1}-\frac{1}{2}\big) \rho c_s^2\Delta t,\alpha\neq \beta,\ \ \mu= \frac{1}{2}\big(s_{s\alpha}^{-1}-\frac{1}{2}\big) \rho(c_\alpha^2- c_s^2)\Delta t,
\end{equation}
\begin{equation}
\mu_{b}= \frac{1}{d}\big(s_{b\alpha }^{-1}-\frac{1}{2}\big) \rho(c_\alpha^2- c_s^2)\Delta t.
\end{equation}
\end{subequations}

It should be noted that due to its inherent characteristics, the present RMRT-LB model can correctly recover the generalized NSEs [Eq. (\ref{eq:4-24-3})] and the special case, i.e., Eq. (\ref{eq:NSE}). However, the present RMRT-LB method is only suitable for the incompressible or weakly compressible fluid flows since the third-order term in Mach number has been omitted, as discussed above.
Finally, we also present a local scheme to calculate shear stress or strain rate tensor in the framework of LB method \cite{Chai2012,Kruger2010,Yong2012}. From Eqs. (\ref{eq:4-23}), (\ref{eq:4-28}) and (\ref{eq:4-29}), one can obtain different approximate formulas of generalized shear stress
\begin{eqnarray} \label{eq:4-32-0}
\mathbf{\tau} &=& -(\mathbf{I}-\mathbf{S}_2/2) \sum_k \big[\mathbf{c}_{k}\mathbf{c}_{k}f_k^{ne} +\frac{\Delta t}{2} (\mathbf{u\bar{F}}+\mathbf{\bar{F}u}+(c_s^2\mathbf{I}-\mathbf{uu})\bar{S})\big]+O(\Delta t^2+\Delta t Ma^{3})\nonumber\\ &=& -(\mathbf{I}-\mathbf{S}_2/2) \sum_k \big[\mathbf{c}_{k}\mathbf{c}_{k}f_k^{ne} +\frac{\Delta t}{2} (\mathbf{u\bar{F}}+\mathbf{\bar{F}u}+c_s^2\bar{S}\mathbf{I})\big]+O(\Delta t^2+\Delta t Ma^{3})\nonumber\\
                   &=& -(\mathbf{I}-\mathbf{S}_2/2) \sum_k \big[\mathbf{c}_{k}\mathbf{c}_{k}f_k^{ne} +\frac{\Delta t}{2} c_s^2\bar{S}\mathbf{I}\big]+O(\Delta t^2+\Delta t Ma^{2}),
\end{eqnarray}
where $\mathbf{uu}\bar{S}=O(Ma^{3})$ and $\mathbf{u\bar{F}}+\mathbf{\bar{F}u}=O(Ma^2)$ are used.

For incompressible fluid flows, the term $O(\Delta t^2+\Delta t Ma^{3})$ in Eq. (\ref{eq:4-32-0}) can be neglected, then the following local scheme for shear stress with a second-order accuracy in time can be obtained from Eq. (\ref{eq:4-32-0})
\begin{eqnarray} \label{eq:4-32}
\mathbf{K}_\mu (\nabla\mathbf{u}+(\nabla\mathbf{u})^{T}) = -(\mathbf{I}-\mathbf{S}_2/2) \sum_k \big[\mathbf{c}_{k}\mathbf{c}_{k}f_k^{ne} +\frac{\Delta t}{2} (\mathbf{u\bar{F}}+\mathbf{\bar{F}u}+c_s^2\bar{S}\mathbf{I})\big],
\end{eqnarray}
where $\mathbf{S}_2$ is a diagonal matrix with $\mathbf{S}_{2,\alpha\alpha}=s_{s\alpha}$,  and $\mathbf{S}_{2,\alpha\beta}=s_{\alpha\beta}$ ($\alpha\neq \beta$).

Based on Eq. (\ref{eq:4-32}), one can also give an expression of strain rate tensor with a second-order accuracy in time,
\begin{eqnarray} \label{eq:4-33}
\frac{\nabla\mathbf{u}+(\nabla\mathbf{u})^{T}}{2} = -(\mathbf{K}_\mu^{-1}/2)(\mathbf{I}-\mathbf{S}_2/2) \sum_k \big[\mathbf{c}_{k}\mathbf{c}_{k}f_k^{ne} +\frac{\Delta t}{2} (\mathbf{u\bar{F}}+\mathbf{\bar{F}u}+c_s^2\bar{S}\mathbf{I})\big].
\end{eqnarray}
It should be noted that if the term $\mathbf{u\bar{F}}+\mathbf{\bar{F}u}$ in Eqs. (\ref{eq:4-26}), (\ref{eq:4-32}) and (\ref{eq:4-33}) are neglected, the much simpler formulas can be obtained with the truncation error $O(\Delta t^2+\Delta t Ma^{2})$.


In addition, from Eqs. (\ref{eq:4-32-0}) and (\ref{eq:4-32}) we obtain
\begin{eqnarray} \label{eq:4-34}
\sum_k \mathbf{c}_{k}\mathbf{c}_{k}f_k^{ne}&=& - (\mathbf{I}-\mathbf{S}_2/2)^{-1}  \mathbf{\tau}  -\frac{\Delta t}{2} (\mathbf{u\bar{F}}+\mathbf{\bar{F}u}+c_s^2\bar{S}\mathbf{I})+O(\Delta t^2+\Delta t Ma^{3})\nonumber\\
                                           &=& - (\mathbf{I}-\mathbf{S}_2/2)^{-1}  \mathbf{\tau}  -\frac{\Delta t}{2} c_s^2\bar{S}\mathbf{I}+O(\Delta t^2+\Delta t Ma^{2}),
\end{eqnarray}
\begin{eqnarray} \label{eq:4-35}
\sum_k \mathbf{c}_{k}\mathbf{c}_{k}f_k^{ne}&=& - (\mathbf{I}-\mathbf{S}_2/2)^{-1}  \mathbf{K}_\mu (\nabla\mathbf{u}+(\nabla\mathbf{u})^{T})  -\frac{\Delta t}{2} (\mathbf{u\bar{F}}+\mathbf{\bar{F}u}+c_s^2\bar{S}\mathbf{I})+O(\Delta t^2+\Delta t Ma^{3})\nonumber\\
                                           &=& - (\mathbf{I}-\mathbf{S}_2/2)^{-1}  \mathbf{K}_\mu (\nabla\mathbf{u}+(\nabla\mathbf{u})^{T})  -\frac{\Delta t}{2} c_s^2\bar{S}\mathbf{I}+O(\Delta t^2+\Delta t Ma^{2}).
\end{eqnarray}

Let $\mathbf{m}_2^{ne}=\sum_k \mathbf{c}_{k}\mathbf{c}_{k}f_k^{ne}$, then utilizing Eqs. (\ref{eq:4-18-1}), (\ref{eq:4-34}) and (\ref{eq:4-35}), one can derive a useful formula to approximate $f_j^{ne}$,
\begin{eqnarray}\label{eq:fne}
f_j^{ne} & = & g_j^{eq}(0, \mathbf{0}, \mathbf{m}_2^{ne})   =  \omega_j (\textbf{c}_j \textbf{c}_j -\mathbf{\Delta}^{(2)} ):\mathbf{\tilde{\mathbf{m}}_2^{ne}}\nonumber\\
         & = & \omega_j  \left[\frac{\mathbf{m}_{2,\alpha\alpha}^{ne}(c_{j\alpha}^2 - c_s^2)}{c_s^2(c_{\alpha}^2-c_s^2)}+\frac{\mathbf{m}_{2,\alpha\bar{\alpha}}^{ne}(c_{j\alpha}c_{j\bar{\alpha}})}{2c_s^4}\right],
\end{eqnarray}
which is an extension of the formula given by Guo et al. in Ref. \cite{GuoZhao2003} when $\mathbf{\Lambda}=\mathbf{S}=\mathbf{I}/\tau, \mathbf{K}_\mu=\mu\mathbf{I}, c_\alpha=c$ and $c_s^2=c^2/3$. We would like to point out that this formula can be used for the initialization of $f_j$.

\subsection{The equilibrium, auxiliary and source distribution functions of RMRT-LB method for NSEs}

From above analysis, one can clearly observe that to recover the macroscopic NSEs (\ref{eq:NSE}) from RMRT-LB method (\ref{eq:2-1}), the equilibrium, auxiliary and source distribution functions should satisfy some necessary requirements, as depicted by Eq. (\ref{eq:4-18}). For a rD$d$Q$q$ lattice model, the explicit expressions of $f_{j}^{eq}$, $G_j$ and $F_j$ with $c_{s \alpha}=c_s$ for all $\alpha$ can be given by
\begin{eqnarray}\label{eq:NSEfeq}
f_j^{eq} & = & g_j^{eq}(\rho, \rho\mathbf{u}, \rho\mathbf{uu}) \nonumber\\
         & = & \omega_j \rho \left[1+\frac{\textbf{c}_{j\alpha}{\mathbf{u}_{\alpha}}}{c_s^2}
                   +\frac{u_{\alpha}^2(c_{j\alpha}^2 - c_s^2)}{c_s^2(c_{\alpha}^2-c_s^2)}+\frac{u_{\alpha}u_{\bar{\alpha}}(c_{j\alpha}c_{j\bar{\alpha}})}{2c_s^4}\right],
\end{eqnarray}
\begin{equation}\label{eq:NSEGeq}
G_j=g_j^{eq}(0, \mathbf{0}, \mathbf{M}_{2G})=\omega_j \left[\frac{\mathbf{M}_{2G,\alpha\alpha}(c_{j\alpha}^2 - c_s^2)}{c_s^2(c_{\alpha}^2-c_s^2)}+\frac{\mathbf{M}_{2G,\alpha\bar{\alpha}}(c_{j\alpha}c_{j\bar{\alpha}})}{2c_s^4}\right],
\end{equation}
\begin{equation}\label{eq:NSEFeq}
F_j=g_j^{eq}(\bar{S}, \mathbf{\bar{F}}, -c_s^2\bar{S}\mathbf{I})=\omega_j \left[\bar{S}+\frac{\textbf{c}_{j\alpha} {\mathbf{\mathbf{\bar{F}}}_{\alpha}}}{c_s^2}
                   -\bar{S}\frac{e_{\alpha}(c_{j\alpha}^2 - c_s^2)}{c_{\alpha}^2-c_s^2}\right],
\end{equation}
where $\mathbf{M}_{2G}$ is determined by Eq. (\ref{eq:4-26}), which can also be computed by
\begin{subequations}\label{eq:M2GComput}
\begin{equation}
\mathbf{M}_{2G}  = \big(\mathbf{I}-\mathbf{S}_2 /2\big)(\mathbf{u}\bar{\mathbf{F}}+\bar{\mathbf{F}}\mathbf{u}+c_{s}^{2}\bar{S}\mathbf{I}),
\end{equation}
\begin{equation}
\mathbf{M}_{2G}  = c_{s}^{2}\bar{S}\big(\mathbf{I}-\mathbf{S}_2 /2\big),
\end{equation}
\end{subequations}
where $\mathbf{uu}\bar{S}=O(Ma^{3})$ and $\mathbf{u\bar{F}}+\mathbf{\bar{F}u}=O(Ma^2)$ have been used to derive Eq. (\ref{eq:M2GComput}a) and Eq. (\ref{eq:M2GComput}b), respectively.

From above discussion, one can find that the NSEs and NCDE can be recovered correctly from the RMRT-LB method through the DTE method, and in other words, the present RMRT-LB method not only can be used to study fluid flow problems governed by the NSEs, but also can be adopted to investigate the heat and mass transport problems depicted by the coupled NCDE(s) and NSEs.

\textbf{Remark} \uppercase\expandafter{\romannumeral1}: The second order moment of $f_j^{ne}$, i.e. $\langle \mathbf{EE} \rangle\mathbf{f}^{ne}=\sum_k \mathbf{c}_k \mathbf{c}_k f_k^{ne}$ in Eq. (\ref{eq:4-32}) or (\ref{eq:4-33}) can be computed by its counterpart ($\langle \mathbf{EE} \rangle\mathbf{f}^{ne}=\mathbf{m}_{\langle \mathbf{EE} \rangle}^{ne}=\mathbf{m}_{\langle \mathbf{EE} \rangle}-\mathbf{m}_{\langle \mathbf{EE} \rangle}^{eq}$) in the moment space (see Eq. (\ref{eq:5-3}) or (\ref{eq:5-4})).

\textbf{Remark} \uppercase\expandafter{\romannumeral2}: It can be seen from Eq. (\ref{eq:4-31}) that the RMRT-LB model on rectangular or standard lattice with $c_s^2\neq c^2/3 $ does not have the versions of the SRT-LB model and TRT-LB model ( \cite{Ginzburg2005a}), since there are at least two different relaxation factors related to the second moment of $f_j^{eq}$ (or to the dynamic viscosity $\mu$).  Keep this in mind, we can find that the rectangular MRT-LBM model in Ref. \cite{Bouzidi2001} has only one relaxation factor $s_{\nu}$ related to the viscosity $\nu$, thus it cannot recover the NSEs (\ref{eq:NSE}) correctly.

\textbf{Remark} \uppercase\expandafter{\romannumeral3}: Even for the standard lattice, if $c_s^2\neq c^2/3 $, we can also obtain the new forms of  $f_j^{eq}, G_j$ and $F_j$ defined in Eqs. (\ref{eq:NSEfeq}), (\ref{eq:NSEGeq}) and (\ref{eq:NSEFeq}) for NSEs, which gives to a class of
MRT-LB models with the sound speed $c_s^2$ as an adjustable parameter.

\textbf{Remark} \uppercase\expandafter{\romannumeral4}: Inserting $\rho=\rho_0+\delta \rho$ with $\delta \rho=O(Ma^2)$ in Eq. (\ref{eq:NSEfeq}), and omitting the terms of $O(Ma^3)$, one can obtain the previous LB model  \cite{HeLuo1997} from the present RMRT-LB model with $f_j^{eq}=g_j^{eq}(\rho, \rho_0\mathbf{u}, \rho_0\mathbf{uu})$.

\textbf{Remark} \uppercase\expandafter{\romannumeral5}: Under the assumption of low Mach number, we can take $\mathbf{a}=0$ or $s_{b\alpha}^{-1}=s_{s\alpha}^{-1}$ for all $\alpha$ in Eq. (\ref{eq:4-27-1}), thus $\mathbf{S}_2^{-1}$ is simplified as a diagonal matrix.

\section{Structure of collision matrix and some special cases of RMRT-LB method}

As we known, the MRT-LB model can also be analyzed and implemented in moment space. When the discrete velocity set $V_q=\{\mathbf{c}_j, 0\leq j\leq q-1\}$ or the rD$d$Q$q$ lattice model is given, we can construct different forms of collision matrix $\mathbf{\Lambda}$ in the RMRT-LB method (\ref{eq:2-1}). In general, the commonly used collision matrix $\mathbf{\Lambda}$ which satisfies the constraints in Eq. (\ref{eq:4-9}) or (\ref{eq:4-18}) has the following form \cite{Chai2020},
\begin{subequations}\label{eq:5-1}
\begin{equation}
\mathbf{\Lambda}=\mathbf{M}^{-1}\mathbf{SM},
\end{equation}
\begin{equation}
\mathbf{S}=(\mathbf{S}_{kj}),\ \ \mathbf{S}_{kj}=0\ (k<j),\ \ \mathbf{S}_{kk}=\mathbf{S}_{k},
\end{equation}
\end{subequations}
where $\mathbf{M}$ is an invertible transformation matrix, whose rows are made up of discrete velocities in $V_q$, $\mathbf{S}$ is a block-lower-triangle matrix, $\mathbf{S}_k \in R^{n_k\times n_k}$ is a relaxation matrix corresponding to the $k$-th ($0\leq k\leq m$) order moment of discrete velocity.

Given a transformation matrix $\mathbf{M}$, let $\mathbf{f}=(f_0,f_1,\cdots,f_{q-1})^{T}$, $\mathbf{f}^{eq}=(f_0^{eq},f_1^{eq},\cdots,f_{q-1}^{eq})^{T}$, then one can execute the transformation between velocity space and moment space
\begin{equation}\label{eq:5-3}
\mathbf{m}=\mathbf{M} \mathbf{f}, \mathbf{m}^{eq}= \mathbf{M}\mathbf{f}^{eq},
\end{equation}
or equivalently,
\begin{equation}\label{eq:5-4}
\mathbf{f}=\mathbf{M}^{-1} \mathbf{m}, \mathbf{f}^{eq}= \mathbf{M}^{-1}\mathbf{m}^{eq}.
\end{equation}

Using Eqs. (\ref{eq:5-1}), (\ref{eq:5-3}) and (\ref{eq:5-4}), the evolution equation of RMRT-LB method (\ref{eq:2-2}) can be written in a matrix form as follows.
\begin{subequations}\label{eq:2-2-1}
\begin{equation}
\textbf{Collison in moment space:}\ \  \tilde{\mathbf{m}}(\mathbf{x}, t)=\mathbf{m}(\mathbf{x}, t)-\mathbf{S} \mathbf{m}^{ne}(\mathbf{x}, t)+\Delta t \big[\mathbf{m}_G(\mathbf{x},t)+\mathbf{m}_F(\mathbf{x}, t)+\frac{\Delta t}{2}\tilde{\mathbf{D}} \mathbf{m}_F(\mathbf{x}, t)\big],  \\
\end{equation}
\begin{equation}
\textbf{Propagation in velocity space:}\ \  \mathbf{f}(\mathbf{x}+\mathbf{c}_j \Delta t,t+\Delta t)=\mathbf{M}^{-1}\tilde{\mathbf{m}}(\mathbf{x}, t), \  \ \ \ \ \ \ \ \ \ \  \ \ \ \ \ \ \ \ \ \ \ \ \ \ \ \ \ \ \ \ \ \ \ \ \ \ \ \
\end{equation}
\end{subequations}
where $\mathbf{f}(\mathbf{x}+\mathbf{c}_j \Delta t,t+\Delta t)=(f_0(\mathbf{x}+\mathbf{c}_0 \Delta t,t+\Delta t), f_1(\mathbf{x}+\mathbf{c}_1 \Delta t,t+\Delta t), \cdots, f_{q-1}(\mathbf{x}+\mathbf{c}_{q-1} \Delta t,t+\Delta t))^{T}$, $\mathbf{m}^{ne}=\mathbf{m}-\mathbf{m}^{eq}, \mathbf{m}_G=\mathbf{MG}, \mathbf{m}_F=\mathbf{MF}, \mathbf{G}=(G_0,G_1,\cdots, G_{q-1})^{T}, \mathbf{F}=(F_0,F_1,\cdots,F_{q-1})^{T}, \tilde{\mathbf{D}}=\mathbf{M}^{-1} \mathbf{diag}(\bar{D}_j)\mathbf{M}$.
On the other hand, one can also implement the RMRT-LB method (\ref{eq:2-2}) in the velocity space,
\begin{subequations}\label{eq:2-2-2}
\begin{equation}
\textbf{Collison:}\ \  \tilde{\mathbf{f}}(\mathbf{x}, t)=\mathbf{M}^{-1}(\mathbf{m}(\mathbf{x}, t)-\mathbf{S} \mathbf{m}^{ne}(\mathbf{x}, t))+\Delta t \big[\mathbf{G}(\mathbf{x},t)+\mathbf{F}(\mathbf{x}, t)+\frac{\Delta t}{2} \mathbf{diag}(\bar{D}_j)\mathbf{F}(\mathbf{x}, t)\big],  \\
\end{equation}
\begin{equation}
\textbf{Propagation:}\ \  \mathbf{f}(\mathbf{x}+\mathbf{c}_j \Delta t,t+\Delta t)=\tilde{\mathbf{f}}(\mathbf{x}, t). \  \ \ \ \ \ \ \ \ \ \ \ \ \ \ \ \ \ \ \ \ \ \ \ \ \ \ \ \ \ \ \ \ \ \ \ \ \ \ \ \ \ \ \ \ \ \ \ \ \ \ \ \ \ \ \ \ \ \ \ \ \ \ \ \ \ \ \ \ \ \ \ \ \
\end{equation}
\end{subequations}

 Similarly, the RMRT-LB method (\ref{eq:2-3}) can also be conducted as
\begin{equation}\label{eq:2-2-3}
  \mathbf{\bar{f}}(\mathbf{x}+\mathbf{c}_j \Delta t,t+\Delta t)=\mathbf{M}^{-1}\big[\mathbf{\bar{m}}(\mathbf{x}, t)-\mathbf{S} \mathbf{\bar{m}}^{ne}(\mathbf{x}, t)+\Delta t (\mathbf{m}_G(\mathbf{x},t)+(\mathbf{I-S/2})\mathbf{m}_F(\mathbf{x}, t))\big],
\end{equation}
where $\mathbf{\bar{f}}=\mathbf{f}-\Delta t \mathbf{F}/2, \mathbf{\bar{m}}=\mathbf{M} \mathbf{\bar{f}}, \mathbf{\bar{m}}^{ne}=\mathbf{\bar{m}}-\mathbf{m}^{eq}$.

From Eqs. (\ref{eq:2-1}), (\ref{eq:2-2}), (\ref{eq:2-2-1}), (\ref{eq:2-2-2}), and (\ref{eq:2-2-3}), we can see that there are two ways to analyze and implement the RMRT-LB method: the way in velocity space and that in moment space. To recover the NSEs (\ref{eq:NSE}) or NCDE (\ref{eq:NCDE}) from the RMRT-LB method, one only needs the basic constraints on the DFs $f_j^{eq}, G_j, F_j$ and the collision matrix $\mathbf{\Lambda}$ given in Eq. (\ref{eq:4-18}) or (\ref{eq:4-9}). Therefore, using the first way we can analyze the LB models under a unified framework, as shown in Ref. \cite{Chai2020}. Compared to the first way, the second one depends on the choice of rD$d$Q$q$ lattice model and the form of the transformation matrix $\mathbf{M}$, and additionally, the equilibrium moment $\mathbf{m}^{eq}$ also needs to be given properly, which leads to the fact that the modeling and analysis of the MRT-LB models are usually limited to the specified space dimension and lattice structure. This may be one of the main reasons why the RMRT-LB model has not been developed well for a long time. However, the analysis method in moment space enables us to design the equilibrium moment $\mathbf{m}^{eq}$ (e.g. the inverse design of LB model as in Refs. \cite{PengGuoWang2019}) more flexibly by selecting the appropriate transformation matrix $\mathbf{M}$, and high-order moments which do not influence the recovery of the macroscopic equation. If necessary, one can obtain the corresponding EDF $f_j^{eq}$ by taking the inverse transformation (\ref{eq:5-4}). It should be noted in this case, the obtained $f_j^{eq}$ may not have the form as Eq. (\ref{eq:4-14}) or (\ref{eq:NSEfeq}).

Actually, there are still some key questions in the RMRT-LB method, for example, what are the relationships among different forms of collision matrices? How do we choose a right one? Thanks to the form of collision matrix in Eq. (\ref{eq:5-1}), one can answer the first question, while the second one needs to be further studied.

It is worth noting that the collision matrices in the commonly used MRT-LB method are of the form in Eq. (\ref{eq:5-1})  \cite{Chai2020}, including the raw (natural) moment, cascaded (central-moment), Hermite-moment, and central-Hermite-moment LB models  \cite{Geier2006, Coreixas2019} with orthogonal and nonorthogonal moment basis vectors. In fact, any two of the transformation matrices $\mathbf{M}$ and $\mathbf{\bar{M}}$ in these MRT-LB models have the following relation,
\begin{equation}\label{eq:5-1-1}
\mathbf{\bar{M}}=\mathbf{NM},
\end{equation}
and one can write the collision matrices with the form in Eq. (\ref{eq:5-1}),
\begin{equation}\label{eq:5-1-2}
\mathbf{\Lambda}=\mathbf{\bar{M}}^{-1}\mathbf{\bar{S}}\mathbf{\bar{M}}=(\mathbf{NM})^{-1}\bar{\mathbf{S}}(\mathbf{NM})=\mathbf{M}^{-1}\mathbf{S}\mathbf{M},
\end{equation}
where $\mathbf{N}$ is a lower-triangle matrix, $\mathbf{S}=\mathbf{N}^{-1}\bar{\mathbf{S}}\mathbf{N}$. Therefore, the existing MRT-LB models mentioned above can be considered as the special cases of the present RMRT-LB method. Furthermore, the present RMRT-LB method includes three classes of MRT-LB models:

(1) The popular MRT-LB models on standard D$d$Q$q$ lattice  with $c_s^2=c^2/3$ \cite{Chai2020}.

(2) The MRT-LB models on standard D$d$Q$q$ lattice with $c_s^2\neq c^2/3$.

(3) The MRT-LB models on the rectangular rD$d$Q$q$ lattice.

Due to the construction of relaxation matrix $\mathbf{S}$ in Eq. (\ref{eq:5-1}b), and based on Eqs. (\ref{eq:5-1-1}) and (\ref{eq:5-1-2}), we only use the transformation matrix $\mathbf{M}$ related to the raw or natural moments in the analysis and implementation of the RMRT-LB model,
\begin{equation}\label{eq:5-2}
\mathbf{m}_{nmp}=\sum_j c_{jx}^m c_{jy}^n c_{jz}^p f_j, \mathbf{m}_{nmp}^{eq}=\sum_j c_{jx}^m c_{jy}^n c_{jz}^p f_j^{eq}, m,n,p \in \{0,1,2\},
\end{equation}
where $\mathbf{m}_{mn}=\mathbf{m}_{mn0}$ and $\mathbf{m}_{mn}^{eq}=\mathbf{m}_{mn0}^{eq}$ for the two-dimensional case. The transformation matrix $\mathbf{M}$ related to the natural moments in Eq. (\ref{eq:5-2}) are composed of $\{c_{jx}^m c_{jy}^n c_{jz}^p, 0\leq j\leq q-1\}$.

Denote $\mathbf{M}$ with the following form,
\begin{equation}\label{eq:5-5}
\mathbf{M}=(\mathbf{M}_0^T,\ \mathbf{M}_1^T,\ \mathbf{M}_2^T,\ \cdots,\ \mathbf{M}_m^T)^T, \ \mathbf{M}_k \in R^{n_k\times q},\ 0\leq k\leq m,
\end{equation}
then from Eq. (\ref{eq:5-1}) we have
\begin{equation}\label{eq:5-6}
\mathbf{M}_k\mathbf{\Lambda}=\sum_{j=0}^{k}\mathbf{S}_{kj} \mathbf{M}_j,\  0\leq k\leq m,
\end{equation}
with the rows corresponding to the zero-th, first and second-order moments,
\begin{subequations}\label{eq:5-7}
\begin{equation}
\mathbf{M}_0=\mathbf{e}=(1,\ 1,\ \cdots,\ 1),\ \ \mathbf{M}_1=\mathbf{E}=(\mathbf{c}_{0},\ \mathbf{c}_{1},\ \cdots,\ \mathbf{c}_{q-1}),\ \mathbf{M}_2=\left(
\begin{array}{c}
\mathbf{M}_{2}^{(1)}\\
\mathbf{M}_{2}^{(2)}\\
\end{array}
\right),
\end{equation}
\begin{equation}
S_0 \in R,\ \ \mathbf{S}_1 \in R^{d\times d},\ \mathbf{S}_2  \in R^{\bar{d}\times \bar{d}},\ \bar{d}=d(d+1)/2,
\end{equation}
\end{subequations}
where $\mathbf{M}_{2}^{(1)}=(\mathbf{c}_{0\alpha}\mathbf{c}_{0\alpha},\ \mathbf{c}_{1\alpha}\mathbf{c}_{1\alpha},\ \cdots,\ \mathbf{c}_{q-1\alpha}\mathbf{c}_{q-1\alpha})$ and  $\mathbf{M}_{2}^{(2)}=(\mathbf{c}_{0\alpha}\mathbf{c}_{0\beta},\ \mathbf{c}_{1\alpha}\mathbf{c}_{1\beta},\ \cdots,\ \mathbf{c}_{q-1\alpha}\mathbf{c}_{q-1\beta})_{\alpha<\beta}$.
According to Eqs. (\ref{eq:4-27}) and (\ref{eq:4-27-1}), we can determine the relaxation matrix $\mathbf{S}_2$ as
\begin{equation}\label{eq:5-8}
\mathbf{S}_2=
\left(
\begin{array}{cc}
    \bar{\mathbf{S}}_2^{(1)} & 0  \\
    0 &  \bar{\mathbf{S}}_2^{(2)}  \\
\end{array}
\right),
\end{equation}
with
\begin{equation}\label{eq:5-8-1}
\bar{\mathbf{S}}_2^{(1)}=\big(\mathbf{S}_2^{(1)}\big)^{-1}=\mathbf{A}-\frac{\mathbf{A}\bar{\mathbf{a}}\mathbf{b}^{T}\mathbf{A}}{1+\mathbf{b}^{T}\mathbf{A}\bar{\mathbf{a}}}, \bar{\mathbf{S}}_2^{(2)}=\big(\mathbf{S}_2^{(2)}\big)^{-1}=\mathbf{diag}(s_{\alpha\beta})_{\alpha < \beta},
\end{equation}
where $\mathbf{A}=\mathbf{diag}(s_{s\alpha}), \bar{\mathbf{a}}=\mathbf{a}/d$, $\mathbf{a}$ and $\mathbf{b}$ are defined in Eq. (\ref{eq:4-27-1}). Note that here the elements $\{s_{\alpha\beta}\}_{\alpha > \beta}$ are not contained in $\bar{\mathbf{S}}_2$ (or $\big(\mathbf{S}_2^{(2)}\big)^{-1}$) since $s_{\alpha\beta}=s_{\beta\alpha}$ for $\alpha\neq\beta$.

For a particular case $\mathbf{S}_{kj}=0 \ (k>j)$, one can obtain
\begin{equation}
\mathbf{S}=\mathbf{diag}(\mathbf{S}_{0},\ \mathbf{S}_{1},\ \cdots,\ \mathbf{S}_{m}),
\end{equation}
and Eq. (\ref{eq:5-6}) would reduce to
\begin{equation}
\mathbf{M}_k\mathbf{\Lambda}=\mathbf{S}_{k} \mathbf{M}_k,\  0\leq k\leq m,
\end{equation}
which can be used to derive the commonly used MRT-LB model  \cite{Guo2013,Kruger2017}.

Note that the matrix $\mathbf{M}$ on rD$d$Q$q$ lattice in Eq. (\ref{eq:5-2}) can be expressed as $\mathbf{M}=\mathbf{D}\mathbf{\tilde{M}}$ with $\mathbf{D}$ being a diagonal matrix.
For the rD2Q9 lattice, $\mathbf{\tilde{M}}$ and $\mathbf{D}$ are given by
\begin{subequations}\label{eq:Md2q9}
\begin{equation}
\mathbf{\tilde{M}}=\left(
\begin{array}{ccccccccc}
    1 & 1  & 1 &  1 &  1 &  1 &  1 &  1 &  1 \\
    0 & 1  & 0 & -1 &  0 &  1 & -1 & -1 &  1 \\
    0 & 0  & 1 &  0 & -1 &  1 &  1 & -1 & -1 \\
    0 & 1  & 0 &  1 &  0 &  1 &  1 &  1 &  1 \\
    0 & 0  & 1 &  0 &  1 &  1 &  1 &  1 &  1 \\
    0 & 0  & 0 &  0 &  0 &  1 & -1 &  1 & -1 \\
    0 & 0  & 0 &  0 &  0 &  1 & -1 & -1 &  1 \\
    0 & 0  & 0 &  0 &  0 &  1 &  1 & -1 & -1 \\
    0 & 0  & 0 &  0 &  0 &  1 &  1 &  1 &  1 \\
\end{array}
\right),
\end{equation}
\begin{equation}
\mathbf{D}=\mathbf{diag}(1, c_1, c_2, c_1^2, c_2^2, c_1 c_2, c_1 c_2^2, c_1^2 c_2, c_1^2 c_2 ^2),
\end{equation}
\end{subequations}
and
\begin{equation}\label{eq:meqd2q9}
\mathbf{m}^{eq}=\mathbf{Mf}^{eq}=\rho \big(1, u, v, c_s^2 +u^2, c_s^2+v^2, u v, c_s^2 u, c_s^2 v, c_s^2(u^2+v^2+c_s^2) \big)^{T},
\end{equation}
where $f_j^{eq}$ is defined by Eq. (\ref{eq:NSEfeq}), while for the He-Luo version [17] with $f_j^{eq,HL}=g_j^{eq}(\rho, \rho_0\mathbf{u}, \rho_0\mathbf{uu})$, we have
\begin{eqnarray}\label{eq:meqd2q9-HL}
\mathbf{m}^{eq,HL} & = & \mathbf{Mf}^{eq,HL}\nonumber\\
                   & = & \big(\rho, \rho_0 u, \rho_0 v, c_s^2\rho +\rho_0 u^2, c_s^2\rho+\rho_0 v^2,\rho_0 u v, c_s^2 \rho_0 u, c_s^2 \rho_0 v, c_s^2(\rho_0(u^2+v^2)+c_s^2\rho) \big)^{T}.
\end{eqnarray}

It can also be found that the orthogonal transformation matrices $\mathbf{M}_{ZKA}$ based on the rD2Q9 and rD3Q27 lattices in Ref. \cite{Zecevic2020} have the form of $\mathbf{M}_{ZKA}=\mathbf{NM}$ with $\mathbf{N}$ being a lower-triangular matrix, and $\mathbf{m}_{ZKA}^{eq}$ in Ref. \cite{Zecevic2020} can also be obtained as
\begin{equation}
\mathbf{m}_{ZKA}^{eq}=\mathbf{M}_{ZKA}\mathbf{f}^{eq}=\mathbf{N} \mathbf{m}^{eq},
\end{equation}
with parameters $\gamma_2=\gamma_5=\gamma_7=c_s^4, \gamma_3=\gamma_4=c_s^2, \gamma_6=c_s^6$. And the diagonal elements in the relaxation matrices $\mathbf{S}_{ZKA}$ and $\mathbf{S}$ are the same, thus Eq. (\ref{eq:4-29}) or (\ref{eq:4-31}) are satisfied for the transformation matrices of natural moment and orthogonal one in Ref. \cite{Zecevic2020}. In addition, due to the fact that there is no the orthogonal transformation matrix for rD3Q19, as stated in Ref. \cite{Zecevic2020}, the RMRT-LB model with rD3Q19 lattice cannot be given \cite{Zecevic2020}, while the present RMRT-LB model with rD3Q19 lattice is ready at hand.

The relationship between the matrix $\mathbf{M}$ in Eq. (\ref{eq:5-2}) and that in Ref. \cite{Bouzidi2001} is interesting. Let $c_1=1, c_2=a$ in Eq. (\ref{eq:Md2q9}b), the matrix $\mathbf{M}$ with Eq. (\ref{eq:Md2q9}) becomes

\begin{subequations}\label{eq:Md2q9-1}
\begin{equation}
\mathbf{M}_a=\mathbf{D}_a \tilde{\mathbf{M}}=\left(
\begin{array}{ccccccccc}
    1 & 1  & 1 &  1 &  1 &  1 &  1 &  1 &  1 \\
    0 & 1  & 0 & -1 &  0 &  1 & -1 & -1 &  1 \\
    0 & 0  & a &  0 & -a &  a &  a & -a & -a \\
    0 & 1  & 0 &  1 &  0 &  1 &  1 &  1 &  1 \\
    0 & 0  & a^2 & 0 & a^2 & a^2 & a^2 & a^2 & a^2 \\
    0 & 0  & 0 &  0 &  0 &  a & -a & a & -a \\
    0 & 0  & 0 &  0 &  0 & a^2 & -a^2 & -a^2 & a^2 \\
    0 & 0  & 0 &  0 &  0 &  a &  a & -a & -a \\
    0 & 0  & 0 &  0 &  0 &  a^2 & a^2 & a^2 & a^2 \\
\end{array}
\right),
\end{equation}
\begin{equation}
\mathbf{D}_a=\mathbf{diag}(1, 1, a, 1, a^2, a, a^2, a, a^2),
\end{equation}
\end{subequations}
and the orthogonal transformation matrix used in Ref. \cite{Bouzidi2001} can be written as
\begin{subequations}\label{eq:Md2q9-2}
\begin{equation}
\mathbf{M}_{BHLL}=\mathbf{N}_{BHLL}\mathbf{M}_a=\left(
\begin{array}{ccccccccc}
    1 & 1  & 1 &  1 &  1 &  1 &  1 &  1 &  1 \\
    0 & 1  & 0 & -1 &  0 &  1 & -1 & -1 &  1 \\
    0 & 0  & a &  0 & -a &  a &  a & -a & -a \\
  -2 r_1 & r_2 & r_3 & r_2 & r_3 & r_1 & r_1 & r_1 & r_1 \\
  -2 r_4 & r_5 & r_6 & r_5 & r_6 & r_4 & r_4 & r_4 & r_4 \\
    0 & 0  & 0 &  0 &  0 &  a & -a & a & -a \\
    0 & -2 & 0 & 2 &  0 & 1 & -1 & -1 & 1 \\
    0 & 0  & -2a & 0 & 2a &  a &  a & -a & -a \\
    4 & -2 & -2 & -2 & -2 &  1 & 1 & 1 & 1 \\
\end{array}
\right),
\end{equation}
\begin{equation}
r_1=1+a^2, r_2=1-2a^2, r_3=-2+a^2, r_4=-1+a^2, r_5=2+a^2, r_6=-1-2a^2,
\end{equation}
\end{subequations}
where
\begin{equation}\label{eq:Md2q9-3}
\mathbf{N}_{BHLL}=\left(
\begin{array}{ccccccccc}
    1 & 0  & 0 &  0 &  0 &  0 &  0 &  0 &  0 \\
    0 & 1  & 0 &  0 &  0 &  0 &  0 &  0 &  0 \\
    0 & 0  & 1 &  0 &  0 &  0 &  0 &  0 &  0 \\
  -2 a^2-2 & 0 &  0 & 3 & 3 & 0 & 0 & 0 & 0 \\
  -2 a^2+2 & 0 &  0 & 3 a^2 & -3/a^2 & 0 & 0 & 0 & 0 \\
    0 & 0  & 0 &  0 &  0 &  1 & 0 & 0 & 0 \\
    0 & -2 & 0 & 0 &  0 & 0 & 3/a^2 & 0 & 0 \\
    0 & 0  & -2 & 0 & 0 &  0 &  0 & 3 & 0 \\
    4 & 0 &  0 & -6 & -6/a^2 &  0 & 0 & 0 & 9/a^2 \\
\end{array}
\right).
\end{equation}
The related equilibrium moments are obtained by
\begin{eqnarray}\label{eq:meqd2q9-1}
\mathbf{m}_{BHLL}^{eq}=\mathbf{M}_{BHLL}\mathbf{f}^{eq}=\mathbf{N}_{BHLL}\mathbf{Mf}^{eq}=\rho \big(1, u, v, \bar{e}^{eq}, \bar{p}_{xx}^{eq}, u v, C_1 u/2, C_2 v/2, C_3 \big)^{T},  \nonumber\\
\bar{e}^{eq}=3(u^2+v^2)+2(3c_s^2-r_1), \bar{p}_{xx}^{eq}=r_4(3 r_1 c_s^2-2a^2)/a^2+3(a^2 u^2-v^2/a^2),
\end{eqnarray}
with
\begin{equation}\label{eq:5-9}
C_1=2(3c_s^2 - 2a^2)/a^2, C_2=2(3c_s^2 - 2), C_3=(9 c_s^4 + 9 c_s^2 u^2 + 9c_s^2 v^2 - 6 c_s^2 - 6 v^2)/a^2 - (6 u^2 + 6 c_s^2 - 4).
\end{equation}

Note that $C_1$ and $C_2$ satisfy the following relation,
\begin{equation}\label{eq:5-10}
C_1=(C_2+4(1-a^2))/a^2,
\end{equation}
which is same as that in Ref. \cite{Bouzidi2001}. When $C_1$ and $C_2$ are taken by Eq. (\ref{eq:5-9}), one can find that the equilibrium moments in Eq. (\ref{eq:meqd2q9-1}) are the same as those in Ref. \cite{Bouzidi2001} except the fourth moment $\varepsilon^{eq}$ which does not affect the correct recovery of the NSEs. This implies that for the RMRT-LB model in Ref. \cite{Bouzidi2001}, one can correctly recover the NSEs (\ref{eq:NSE})
 when choosing the relaxation sub-matrix $\bar{\mathbf{S}}_2$ corresponding to $\mathbf{S}_2$ in Eq. (\ref{eq:5-8}). It can be seen from Eq. (\ref{eq:5-8}) or (\ref{eq:4-31}) that in two-dimensional case up to five different relaxation parameters are generally used to determine the dynamic shear viscosity $\mu$ and the dynamic bulk viscosity $\mu_b$, three for $\mu$ and two for $\mu_b$, while at most three relaxation parameters ($s_2,s_8$, and $s_9$) were used in Ref. \cite{Bouzidi2001}, two for $\mu$ and one for $\mu_b$.

\section{Conclusions}

In this work, following our recent work \cite{Chai2020}, we developed a general framework of the RMRT-LB method on the rD$d$Q$q$ lattice for the NCDE and NSEs in a natural and unified way, in which a block-lower-triangular-relaxation matrix is used. At first, the rectangular rD$d$Q$q$ lattice models, the properties of velocity moments and the expression of weight coefficients are discussed, and a general REDF is obtained based on the work in Ref. \cite{LuChaiShi2011}. Then we conducted a detailed DTE analysis on the present RMRT-LB method, and a generalized NSEs with anisotropic viscosity (viscosity tensor) can be correctly recovered, while the analysis of the RMRT-LB method for NCDE is omitted since it is the same as that in Ref. \cite{Chai2020}. It should be noted that if the isotropic viscosity is considered, the popular NSEs will be recovered. Further, the structure of the collision matrix in the RMRT-LB method is analyzed, and we find that some existing MRT-LB models can be viewed as the special cases of the present RMRT-LB method, including the classical MRT-LB model, central-moment LB model, Hermite-moment and central-Hermite-moment LB models. In addition, two new versions of those MRT-LB models with third-order anisotropic moments: one on standard D$d$Q$q$ lattice with $c_s^2\neq c^2/3 $, and the other on rectangular rD$d$Q$q$ lattice where the rectangular MRT-LB model in Ref. \cite{Zecevic2020} is only a special case. 

\section*{Acknowledgements}
This work was financially supported by the National Natural Science
Foundation of China (Grants No. 12072127 and No. 51836003).


\end{document}